\documentstyle[eqsecnum,prb,aps]{revtex}
\def\lsim{\raise1pt\hbox{$<$}\lower3pt\hbox{\llap{$\sim$}}}
\begin{document}
\draft
\twocolumn[\hsize\textwidth\columnwidth\hsize\csname 
@twocolumnfalse\endcsname
\title{Berry's Phase and Quantum Dynamics 
of  Ferromagnetic Solitons}
\author{Hans-Benjamin Braun and Daniel Loss} 
\address{
Department of Physics, Simon Fraser University, 
Burnaby, British Columbia V5A 1S6, Canada
}
\date{Received 31 May, 1995}
\maketitle
\begin{abstract}
%\widetext
%
We study spin parity effects and the quantum 
propagation of solitons (Bloch walls) in 
quasi-one dimensional ferromagnets.
Within a coherent state path integral approach we 
derive a quantum field theory for nonuniform spin configurations. 
The effective action for the soliton
position is shown to contain a gauge potential 
due to the Berry phase and a damping term 
caused by the interaction between soliton and 
spin waves. For temperatures below the anisotropy gap 
this dissipation reduces to a pure soliton mass
renormalization.
The quantum dynamics of the soliton in a periodic lattice or pinning 
potential reveals remarkable consequences of the Berry phase. 
For half-integer spin, destructive interference
between opposite chiralities suppresses nearest neighbor
hopping. Thus the Brillouin zone is halved, and for 
small mixing of the chiralities
the dispersion reveals a surprising dynamical correlation:
Two subsequent band minima belong to different chirality states
of the soliton. 
For integer spin the Berry phase is inoperative and 
a simple tight-binding dispersion is obtained. Finally it 
is shown that  external fields can be used to interpolate 
continuously  between the Bloch wall dispersions 
for half-integer and integer spin. 
\end{abstract}
%\pacs{PACS numbers: 05.40.+j, 75.10.Hk, 75.60.Ch, 75.60.Ej, 75.60.Jp,
%85.70.Li, 87.40.+w, 91.25.Ng}
]
\section{Introduction}
\label{introduction}

Quantum effects in low-dimensional magnetism 
are a fascinating subject which has attracted much interest over the years.
A notable example are antiferromagnetic chains where the
quantum spin (or Berry\cite{berry}) 
phase  leads to remarkable parity effects. It is for integer
spin $S$ only that the groundstate exhibits an excitation 
(or Haldane \cite{haldane})
gap whereas
for half-odd integral $S$ such  gaps are suppressed by
interfering Berry phases.\cite{affleck} Related to this
phenomenon is the suppression of mesoscopic stiffness
fluctuations for $S$ 
half-integral antiferromagnets, whereas such fluctuations grow with
chain size for integer $S$ (similar to universal conductance
fluctuations in mesoscopic metals). \cite{lossmaslov}

Over the last years, the rapid advances in nanostructure technology
\cite{physicstoday} have opened the door to a new class
of magnetic systems: Small single domain particles that display striking 
mesoscopic quantum phenomena\cite{nato} (MQP) such as quantum coherence, 
quantum tunneling, or spin parity effects.
These particles exhibit one or several
directions of minimal anisotropy energy between which 
the spins can tunnel coherently.
Motivated by theoretical predictions for  uniform ferromagnets 
\cite{schilling,vanhemmen,chudi,kim}
and antiferromagnets \cite{barbchud} several experiments
at sub-Kelvin temperatures have 
either shown temperature independent relaxation phenomena 
\cite{paulsen,wernsdorfer,vincent}
or a well-defined resonance \cite{awschalom} (in the 
$ac$ susceptibility)
which scales exponentially with the 
number of spins \cite{awschalomii} in accordance with theory.
\cite{barbchud}
Although these observations have been criticized 
on the basis of dissipation models,
such as the influence of nuclear spins\cite{garg}, 
the experiments on antiferromagnetic ferritin\cite{awschalom}
provide a strong indication
that the spins indeed tunnel at low temperatures.

In subsequent work, it has been shown \cite{LDG,delft,Ldgas,BLii}
that also tunneling  depends on the spin parity via Berry phases, 
and that the magnetization switching 
is allowed only if the total spin 
of the particle is integral but not otherwise. 
Similar results have been found in uniform 
antiferromagnetic particles.\cite{LDG,Ldgas,chudICM,duan} 

While  
such spin parity effects  are sometimes related to 
Kramers' degeneracy, in particular in single domain 
ferromagnets \cite{schilling,vanhemmen,LDG}, they 
typically go  beyond this theorem in rather unexpected ways.
\cite{LDG,BLii} 
It is notably for {\it non-uniform} magnets that such effects
can be quite intriguing as we know since Haldane's work on 
antiferromagnets. \cite{haldane}
On the other hand, there has not been much 
related study on non-uniform ferromagnets, primarily because their 
groundstate is trivial and did not seem to offer much room for 
surprises. However, this is by no means so, and it is one 
of our goals  to show that 
ferromagnets with more than one magnetic domain do exhibit novel
spin parity effects and that these effects can lead to
experimental consequences.

We address the issue of spin parity in the context of MQP, 
although the Berry phase effects
discussed here are of general relevance in low-dimensional magnetism.
We start by considering the coherent quantum propagation
of Bloch walls in the presence of periodic pinning
potentials. Such potentials are naturally provided by the underlying
crystal lattice or some  superlattice structure 
that can be created by periodic deposition of materials with different 
anisotropies.\cite{BL,BLii,klitzing} 
Parenthetically we note that periodic pinning provides a 
much smaller barrier height and tunneling distance 
than one isolated pinning center would do. 
Thus the tunneling probability will be drastically 
enhanced  in this case \cite{BL,BLii} compared to the more 
traditional scenario where experiment 
\cite{barbara,giordano,wernsdorfer} and 
theory \cite{egami,riehemann,stamp,scb,tatara} focus on
wall tunneling out of single pinning centers.

In a collective coordinate description the 
Bloch wall is then seen to behave like a single degree of freedom 
moving in a periodic structure.\cite{BL} 
This in turn results in characteristic
Bloch bands in reciprocal space, where the bandwidth is determined
by the tunneling rate through the potential. 
It is now at this stage where the Berry 
phase enters the wall dynamics 
via an effective  gauge potential that depends on the
{\it chirality}, i.e.
the internal rotation sense of the Bloch wall. For half-integer spins
this gauge potential
induces  a halving of the associated
Brillouin zone. At the same time a remarkable dynamical correlation
occurs: Two subsequent band minima belong to opposite chiralities. 
Thus, the chirality of the wall 
alternates when the system is adiabatically driven  through the 
Brillouin zones by an external magnetic field. 
As we shall argue, this phenomenon
can be experimentally observed if there is a finite 
tunneling probability between the chiralities, a condition which is 
not difficult to meet in real systems. 
Due to the topological nature of the Berry phase these 
results are independent of details such as shape of the
soliton and the pinning potential. Thus we expect that band-halving and 
chirality correlation also occurs in the limit of a
spin-${1\over2}$ chain where the soliton width approches 
one lattice constant.

Besides these spin parity effects,
the bandstructure 
leads to interesting coherence effects in form of Bloch 
oscillations of the wall center. \cite{BL,BLii}  As a result the 
sample magnetization oscillates in response to a {\it static} 
magnetic field, a behavior which is very similar to the $ac$-Josephson
effect in superconductors.

In principle these  results hold for an  arbitrary 
number $N_A$ of coupled ferromagnetic chains.
However, observation 
of MQP becomes increasingly difficult 
with increasing $N_A$ since observability 
requires tunneling exponents (which grow with $N_A$) to be of 
the order of Planck's constant. 
This necessarily limits the size of sample cross sections (but not
their lengths) and restricts considerations to low-dimensional
ferromagnets, most typically of quasi-one dimensional size.
An important consequence of this reduced dimensionality is the fact
that dissipation due to spin waves has a negligible effect 
on the wall dynamics since there is an associated finite
size gap (besides the anisotropy gap) in the spin wave  
spectrum.
It is due to these gaps that at temperatures 
typically below $100 \;{\rm mK}$ the spin waves freeze out 
exponentially fast, and are thus simply irrelevant for dissipation
(they only lead to a minor soliton mass renormalization
as we shall show explicitly.)

To simplify our discussion we consider in the following the 
limit of large hard-axis anisotropy 
\cite{footnote:general,lossbraun} as 
it occurs for instance in an Yttrium-Iron-Garnet (YIG) sample 
of the shape shown in Fig \ref{sample}. 
We can then eliminate the out of easy-plane degree of freedom and  
the spin model  reduces to that of a sine-Gordon 
model plus a gauge term coming from the Berry 
phases.  In a quantum field approach
we introduce collective coordinates, eliminate 
the spin waves and arrive at an effective action for the wall
position.
The spin waves give rise to a non-local term in the action
which can be cast into the well-known Caldeira-Leggett 
form  at low temperatures.
In this way we make contact with phenomenological formulations of 
dissipation \cite{caldleg,weiss} as extensively  discussed in 
the context
of MQP \cite{leggett} and show that the spectral function has
a gap due to anisotropies.
While there has been a number of  work in 
various contexts  related to intrinsic soliton damping
\cite{janak,abyzov,ivanov,bary,neto,stamp}, we believe that
the novel approach presented here is most adequate to the combined
description of wall dynamics and Berry phases and, moreover, 
provides the first complete discussion
of spin wave dissipation, particularly in the context of MQP.
Finally we note that a brief account of part of the results 
presented here has been given before. \cite{BL,BLii}.

The outline of the paper is as follows. 
In Sec. \ref{model} we discuss the derivation 
of the spin  action plus topological phase 
from the Heisenberg model. 
Details of this derivation via coherent spin states
together with a unified
treatment of the Berry phase in different gauges
are given in  Appendix \ref{appendixa}.
In Sec. \ref{blochwalls} we discuss static Bloch wall solutions
and derive the sine-Gordon action plus gauge term 
in Sec. \ref{sinegordon}. To gain confidence
in our approach we first consider the uniform limit and show that
this gauge term reproduces the known spin parity behavior. \cite{LDG}
As a byproduct we also obtain the tunneling prefactor.
In Sec. \ref{dissipation} we discuss the coupling 
between Bloch wall and spin waves and 
show that spin wave dissipation is negligible at low temperatures,
technical details are presented in Appendix \ref{appendixb}. 
In Sec. \ref{interference} we  discuss the influence
of the Berry phase on the Bloch wall dynamics in a periodic
potential, first in the nearly free (Sec. \ref{interference2}) and 
then in the tight binding limit (Sec. \ref{interference3}). 
In both cases it  is shown that for half integral spin the 
Brillouin zone is halved and the chirality alternates.
Experimental implications are given in Sec. \ref{expts}, where
we also give results for the level splitting
due to the tunneling between the two wall chiralities.
Finally in Sec. \ref{extfields} we discuss how the interference
effects are altered by external fields. A note 
regarding the terminology: The terms soliton and Bloch 
(or domain) wall are used interchangeably 
to denote the transition region between domains in ferromagnets.

\section{Model} 
\label{model}
In this section  we derive a continuum field theory to describe the 
quantum dynamics of nonuniform spin configurations in ferromagnets. 
Our starting point is  a microscopic Heisenberg spin-Hamiltonian with
local anisotropies.  The transition amplitude between two
nonuniform spin configurations  is then expressed as 
a coherent state path integral.  The corresponding action 
differs from  the classical micromagnetic expression  by a
total derivative. 
While this term does not affect the classical equations of motion, 
it gives rise to quantum mechanical interference effects
and thus leads in a natural way to the quantization of 
micromagnetics.  Several examples of such interference effects will 
be discussed below in  Sections \ref{sinegordon},\ref{interference}-
\ref{extfields}. 

Ferromagnetic insulators can often be described by a  Heisenberg 
Hamiltonian with anisotropies
\begin{equation}
{\cal H}\!=\!- \tilde J \!\sum_{i,\rho} {\bf S}_i 
\cdot {\bf S}_{i+\rho} -\tilde K_y\! \sum_i (S_i^y)^2 
+\tilde K_z \!\sum_i (S_i^z)^2, 
\label{H}
\end{equation}
where ${\bf S}_i$ denotes the spin operator at  the  lattice site $i$.
For simplicity we assume that the  spins form a simple cubic 
lattice of lattice constant  $a$.  Throughout this 
work we shall use units such that  $\hbar=1$.
The first term on the rhs of
(\ref{H}) is  the exchange interaction between a spin  at the lattice
site $i$ and its nearest neighbors at the lattice sites $i+\rho$.  The
next term is an easy-axis anisotropy  along the  $y$-axis with
anisotropy constant $\tilde K_y>0$. The third term is a hard-axis
anisotropy  of strength $\tilde K_z>0$ which renders the 
$xy$-plane an easy plane.  The spins will thus preferably point 
parallel or antiparallel  to the $y$-axis.

The anisotropies that are used in  (\ref{H}) are  effective 
anisotropies  \cite{prb1} and  may arise from two different 
microscopic mechanisms. One contribution is the 
magnetocrystalline anisotropy 
which is due to the interaction of the magnetic moments 
with their neighboring atoms via spin-orbit interaction.  
Consequently this contribution reflects the symmetry of the 
crystal lattice.  The second contribution  is the 
dipolar interaction between the magnetic moments.
Due to its long range nature 
this contribution depends on the sample shape
and is in general a nonlocal functional of the 
magnetization configuration. It is this magnetostatic 
interaction that gives rise to the existence of domains in 
macroscopic samples.
However, for quasi-one dimensional configurations
this interaction  considerably simplifies and can be 
modeled  by local anisotropies as in (\ref{H}).

Our focus in this work will be on elongated samples as shown 
in Fig. \ref{sample} with transverse dimensions
smaller than the length  scale $a [\tilde J/\tilde
K_z]^{1/2}$. Spin waves running transverse 
to the sample\cite{janak,winter} then exhibit a 
finite size gap  such that they are frozen out at 
low temperatures.\cite{gap}  
This condition is met in most experimental situations
studied so far and thus we shall use a 
quasi-one dimensional model in the following. 
Truly three dimensional samples where all degrees of freedom 
are allowed to be excited are of rather limited 
interest for MQP since their tunneling rates  
{\it and} associated crossover temperatures
(separating classical from quantum regime) are in
general too small to 
be observed.\cite{BLii}

We now turn to the path-integral formulation of the
system described by the Hamiltonian (\ref{H}).
We introduce coherent spin states \cite{klauder} at each lattice site,
defined by ${\bf \Omega}_i\cdot {\bf S}_i 
|{\bf \Omega}_i\rangle=S |{\bf \Omega}_i\rangle$
where  ${\bf \Omega}=(\sin\theta \cos\phi, 
\sin\theta \sin\phi,\cos\theta)$ is a unit vector.
The whole system is then  described by a
product of coherent  states  at  each 
of the $N_L$  lattice sites, i.e.
$| \{{\bf \Omega}\}\rangle=\bigotimes_{i=1}^{N_L}
|{\bf \Omega}_i \rangle$.
Since we are interested in configurations that are varying
slowly compared to the lattice constant, the spin state
can be described by a smoothly varying unit vector field 
${\bf \Omega}(x,\tau)$ depending on the 
coordinate $x$ along the sample and the imaginary 
time $\tau$.  As outlined in Appendix A,
the transition amplitude between 
the two states   $|\{{\bf \Omega}_a\}\rangle$,
$|\{{\bf \Omega}_b\}\rangle$ can then be 
expressed as an (imaginary) time path integral, 
\begin{equation}
\langle \{{\bf \Omega}_b\} |  e^{-\beta {\cal H}} |  
\{{\bf \Omega}_a\}\rangle
=\int \! {\cal D} \phi \; {\cal D} (\cos\theta)
e^{-{\cal S}_E[\phi,\theta] },
\label{pathintintro}
\end{equation}
where the 
integration is over all configurations that satisfy 
the boundary  conditions  ${\bf \Omega}(x,0)={\bf \Omega}_a(x)$,
${\bf \Omega}(x,\beta)={\bf \Omega}_b(x)$ (spatial boundary conditions
will be specified later). 
The Euclidean action is given by 
\begin{equation}
{\cal S}_E =  {\cal S}_{WZ} + \int_0^\beta\! d\tau\; H,
\label{SE}
\end{equation}
with the  Wess-Zumino or Berry phase term 
\begin{equation}
{\cal S}_{WZ} = i {S N_A\over a} 
 \int_0^\beta d\tau  \int_{-L/2}^{L/2} \!dx\;
 \dot \phi \; (1-\cos\theta),
\label{swzintro}
\end{equation}
with $\dot \phi=\partial_\tau \phi$ and 
where a gauge has been chosen with the coherent 
states underlying the path integral (\ref{pathintintro})
expressed in the ``north-pole" parametrization 
(cf. Appendix \ref{appendixa}). Eq. (\ref{swzintro}) has for 
closed trajectories the form of the sum over the Berry phases
of all $N_A L/a$ spins. The  energy is given by 
\begin{eqnarray}
H &=& N_A  \int_{-L/2}^{L/2}  dx 
\Biggl \{  J[ (\partial_x \theta)^2 + \sin^2\theta 
(\partial_x \phi)^2] \nonumber \\
&-&K_y[\sin^2 \theta \sin^2 \phi -1] + K_z\cos^2 \theta \Biggr\}.
\label{h}
\end{eqnarray}
$N_A$ denotes the number of spins in the cross 
sectional area ${\cal A}$ of the sample, and $L$ 
is  the sample length. The parameters 
in (\ref{h}) are related to those  in (\ref{H}) 
via 
\begin{equation}
J=\tilde J S^2 a$,  $K_{y,z}=\tilde  K_{y,z} S^2 /a.
\label{parameters}
\end{equation}
The energy (\ref{h})
is identical to the traditional 
micromagnetic energy expression.\cite{prb1,sloncz,brown,dillon}
$J$  and $K_{y,z}$ can now be related to the micromagnetic
anisotropy and exchange constants
$J=A a^2$, $K_y= K_e a^2$,  $K_z=K_h a^2$.
For an elongated slab as shown in Fig. \ref{sample},
we have $K_e=K_{e,{\rm cryst}}$;
$K_h=K_{h,{\rm cryst}}+ 2\pi M_0^2$, where 
$M_0=g \mu_B S/a^3$ is the saturation magnetization,
and $K_{e,{\rm cryst}}$, $K_{h,{\rm cryst}}$  
describe crystalline anisotropies.
Note, however, that for  other sample geometries 
the demagnetizing term enters in a different form.
For instance, for a cylindrical wire we would have 
$K_e=K_{e,{\rm cryst}} +\pi M_0^2$,
while the hard-axis anisotropy would be of purely 
crystalline origin. [For other examples
see Fig. 8 of Ref. \onlinecite{prb1}].  
The demagnetizing energy is not always important, in particular
for samples with misoriented anisotropy axes
[see p. 15 of Ref. \onlinecite{sloncz}]
the crystalline anisotropies can be  
much larger than $2\pi M_0^2$. 

In saddle point  approximation, $\delta {\cal S}_E=0$, 
and rotating to real time $t=-i\tau$, 
we recover the classical Landau-Lifshitz equations  of motion
\begin{eqnarray}
\sin\!\theta\;\partial_t \phi &=&-\;{a\over S}\;{\delta H\over
\delta \theta} , \nonumber \\ 
\partial_t \theta&=& {a\over S}\; {1\over
\sin\theta}\; {\delta H \over \delta \phi}.
\label{eqofmot}
\end{eqnarray}
These classical equations are not affected
by the total derivative $\dot \phi$ in (\ref{swzintro})
and thus follow from the classical Lagrangian density\cite{brown} 
${\cal L}= - {S N_A\over a}\partial_t \phi \cos\theta + H$.

Note, however, that the $\dot\phi$ term is of 
crucial importance for the quantum dynamics: 
While  the  path integral (\ref{pathintintro}) contains 
higher winding number contributions 
where a path retraces itself, the Wess-Zumino term
enforces quantization by 
destructive interference of paths which do not satisfy the 
condition $S\sum_i A_i = 2\pi n$, 
where $A_i$ is the area enclosed by the 
trajectory of the $i$-th spin on the unit sphere. 
If the $\dot \phi$ term were dropped in (\ref{swzintro})
--- a ``gauge" that has sometimes been used in the literature --- 
the area $A_i$ would be measured with respect to the 
equator and one would have  to impose the additional 
constraint \cite{auerbach} that the paths not 
intersect the ``dateline". This  constraint is 
very difficult to handle within a path integral formalism.
On the other hand, ignoring this constraint would lead to 
a wrong semiclassical
quantization of half-integral spins. Moreover, one would not 
obtain the suppression of tunneling for half-integer 
spins in small ferromagnetic particles \cite{LDG}, in 
clear contradiction to Kramers' theorem which requires that
the ground state not be split.
In Appendix A we show that all these difficulties 
can be avoided if one starts from one single premise ---
the single-valuedness of the coherent states ---
which leads to a restricted set of ``admissible" gauges.

Finally, we remark that if we work in the south-pole 
parametrization of the coherent state
(cf.  \ref{cs2}), the $\dot \phi$-term 
in (\ref{swzintro}) changes sign but, of course, 
all physical effects that will be derived below 
are independent of the  gauge (provided the gauge is
admissible).

\section{Bloch wall configurations}
\label{blochwalls}

There are two energetically degenerate 
spin configurations which minimize the 
energy  (\ref{H}):  Uniform configurations with all spins 
pointing either along the positive or along the negative 
$y$-direction. 
We are now interested in structures
that interpolate between these
two configurations. Due to the 
easy-axis anisotropy in (\ref{H}), this
transition region will have a finite
width and form a Bloch wall (or soliton). 
Such Bloch walls may have various origins
in realistic samples. They can  simply be enforced
by keeping the spins at both sample ends antiparallel
to each other. For certain sample geometries,
their existence can be favored by long range 
magnetostatic interactions which have not been built 
into (\ref{h}). Finally,  in strictly one-dimensional
chains, solitons with  width of one lattice constant
rather than spin waves can form the elementary excitations.
\cite{mikeska,spinhalf}

A static Bloch wall connects the anisotropy minima 
$\phi=\pm {\pi\over 2}$ within the easy plane 
$\theta={\pi\over 2}$ and thus satisfies the 
Euler-Lagrange equations 
\begin{equation}
J\partial_x^2 \phi + K_y \sin\phi \cos\phi =0.
\label{blochEL}
\end{equation}
With the additional condition 
$\partial_x\phi(\pm \infty)=0$,
this can be immediately integrated once
\begin{equation}
{J  \over K_y} (\partial_x \phi)^2 -\cos^2 \phi=0.
\label{encons}
\end{equation}
This equation exhibits the symmetries 
$\phi \to -\phi$ and $\phi \to\phi+\pi$,
which reflect the fact that the 
energy (\ref{H}) is  invariant under  rotations 
by $\pi$ around  each axis in spin space.
Consequently,  there are four different 
Bloch wall solutions of (\ref{encons}) 
(see, e.g., Ref. \onlinecite{sloncz}) 
\begin{equation}
\phi_{QC} (x)= -QC {\pi\over 2} + 2 \arctan e^{Cx/\delta},
\qquad \theta={\pi\over 2},
\label{phiqc}
\end{equation}
of width $\delta=\sqrt{J/K_y}$. In order to distinguish 
the  four different soliton configurations 
we have introduced the ``charge"
$Q={1\over 2} \int\! dx\;\partial_x (\sin\phi)$,
and the ``chirality"
\begin{equation}
C={1\over \pi}\int_{-\infty}^{\infty}\! 
dx \;\partial_x \phi, 
\end{equation}
of a spin configuration. For the Bloch walls (\ref{phiqc}) 
we have $Q,C=\pm 1$, and all four walls have the same energy 
\begin{equation}
E_0=2 J N_A\int_{-\infty}^{\infty}\!dx\;(\partial_x \phi_{QC})^2 = 
4 N_A \sqrt{J K_y}.
\label{wallenergy}
\end{equation}
The definition of the chirality $C$
simply tells us whether the 
angle $\phi$ increases or decreases as we proceed 
in positive $x$-direction along the sample. 
The definition of charge $Q$ is motivated by 
the response of the Bloch wall to an applied magnetic field: 
For an external field along the positive $y$-axis, a Bloch 
wall of positive charge moves along the positive $x$-axis
while a negatively charged wall 
moves in the opposite direction. 
[We recall that the spin is antiparallel to the magnetization.]
Within the present description, 
the spin is allowed to point into an arbitrary
direction on the unit sphere $S^2$. In this
case, only the charge is a topological invariant, 
i.e., for infinite sample length 
field configurations of opposite charge 
cannot be  deformed into each other without 
overcoming an infinite energy  barrier. 
Solitons of different chirality 
(but same charge) can be deformed into each other
via a ``N\'eel-wall" configuration where the
spin at the wall center points along the hard-axis.
It is only in the $XY$-limit of large
hard-axis anisotropy where the configurations 
space of the spins becomes a circle and 
the chirality also becomes a topological invariant.
It will be this limit which shall be considered 
in the next section, but we shall return to the
general case  when we discuss chirality 
tunneling in Sec. \ref{expts}.

\section{Relation to the sine-Gordon model}
\label{sinegordon}

In some materials such as
elongated YIG films (cf. Fig. \ref{sample})
or in garnet crystals with 
misoriented anisotropy axes, the 
hard-axis aniotropy is  much larger than the 
easy-axis anisotropy, typically by a factor of ten or more.
As a consequence,  deviations
away from the easy-plane become energetically
costly and the magnetization will be confined to 
the easy-plane and the 
system can effectively be described in the
easy-plane variable $\phi$ only.

In the limit $K_z \gg K_y$, deviations 
away from the easy-plane are suppressed and 
we can expand \cite{compactness}
\begin{equation}
\theta(x,\tau)=\pi/2 - \vartheta(x,\tau),
\label{smalltheta}
\end{equation}
where $|\vartheta| \ll 1$.
Inserting (\ref{smalltheta}) into the action
(\ref{swzintro}) we obtain up to second order in 
$\vartheta$,
\begin{eqnarray}
{\cal S}_E=N_A \int\!\!dx \, d\tau
\Biggl\{   
&i& {S\over a} \partial_\tau \phi + J (\partial_x\phi)^2
+ K_y \cos^2 \phi
\nonumber \\
 &-&i {S\over a} \vartheta \; \partial_\tau\phi 
+ \vartheta \;{\cal L} \; \vartheta \Biggr\},
\label{Stheta}
\end{eqnarray}
where 
${\cal L}=-J\partial_x^2 -J(\partial_x\phi)^2 
+ K_y \sin^2\phi+ K_z$.  If the  fluctuations
in both $\vartheta$ and $\phi$ have wavelength $\lambda$
larger than the domain wall width,
$\lambda \geq\delta$, the  hard-axis 
anisotropy becomes dominant and ${\cal L}
=K_z+ {\cal O}(K_y/K_z)$. With this approximation,
we insert (\ref{smalltheta}) into 
(\ref{pathintintro}) and 
using  ${\cal D} \cos\theta \simeq {\cal D} \vartheta$
we can perform the Gaussian integrations. 
The transition amplitude can then be expressed as 
a path integral over the azimuthal angle $\phi$ alone
\begin{equation}
\langle\{{\bf \Omega}_b\} |
e^{-\beta {\cal H}} | \{{\bf \Omega}_a\}\rangle
\simeq \int {\cal D} \phi \; e^{-{\cal S}_{SG}[\phi]},
\label{SGpath}
\end{equation}
with the boundary conditions $\phi(x,0)=\phi_a(x),\phi(x,\beta)=\phi_b(x)$.
The action has the following form  
\begin{eqnarray}
{\cal S}_{SG} = N_A \int\!\! dx \, d\tau
\Biggl\{i{S\over a} \partial_\tau\phi 
&&+ J[ {1\over c^2} (\partial_\tau \phi)^2 
+ (\partial_x\phi)^2] \nonumber \\
&&+ K_y \cos^2 \phi \Biggr\}.
\label{SG}
\end{eqnarray}
where we have introduced the asymptotic spin wave velocity,
\begin{equation}
c=(2a/S) \sqrt{J K_z}.
\label{spinvel}
\end{equation}
We thus have arrived at the important result that 
{\it for large hard-axis anisotropy, the dynamics
of a mesoscopic ferromagnet is described
by the sine-Gordon (SG) action plus a topological 
term $iSN_A\int\!{dx\over a}\int\!d\tau\,\dot\phi$.}
While the reduction to the sine-Gordon model 
has been known for some time \cite{enz,chudi}, 
the topological term  has not been identified 
before.  This term is of central importance 
for the quantization of the spin system as we shall see 
below. It is this term that is responsible 
for observable effects such as band halving and chirality 
correlation. 

We can now explicitly verify the consistency 
of our approach. ${\cal S}_{SG}$
has the same long wavelength excitations as
the full magnetic model described by the action
(\ref{SE}). In the latter model, deviations
from the uniform state $\phi={\pi\over 2},\theta={\pi\over2}$
along the easy-axis have the spin wave spectrum 
$\omega_k=2(a/S)([Jk^2+K_y+K_z][Jk^2+K_y])^{1/2}$,
with $k$ the spin-wave wavevector.\cite{spinwaves}
For $K_z\gg K_y$  and $k<\sqrt{K_z/J}$
this reduces to 
$\omega_k=2(a/S)(K_z[Jk^2+K_y])^{1/2}=
c\sqrt{k^2+\delta^{-2}}$ which 
is identical to the  spin wave spectrum 
in the sine-Gordon model (\ref{SG}) around 
$\phi=\pi/2$.
Similarly, the dynamic soliton solutions of 
the spin system (see, e.g. Ref. \onlinecite{breath}) 
which correspond to moving Bloch walls  
have their counterpart in the  SG-model in this limit. 
Even soliton-antisoliton breather solutions 
of the spin system have analogues in  SG
breather solutions.\cite{breathers}  This is  
surprising since in the spin model
breathers exhibit a precession around the 
easy-axis and thus do  not stay 
close to the easy plane as  required 
in (\ref{smalltheta}). The connection between 
the spin model and the SG system is therefore 
more general than the above derivation 
suggests.

\subsection{ Spin Tunneling for $K_z\gg K_y$ }

To illustrate the importance of the topological 
term derived above, we consider the case of 
a uniform spin configuration as, e.g.,
realized in a nanoscale ferromagnetic particle. 
We shall show that the reduced model (\ref{SG}) 
reproduces both the spin parity effect \cite{LDG}
and the tunneling action \cite{chudi} of the full 
magnetic model in the XY--limit. 
In addition we shall also 
evaluate the prefactor of the transition amplitude 
resulting from Gaussian fluctuations around the 
instanton path.

For uniform configurations, 
$\phi=\phi(\tau)$,  the action 
(\ref{SG}) reduces to 
\begin{equation}
\bar{\cal S}_{SG}=N\int d\tau \Bigl\{ iS \partial_\tau
\phi + {J a\over c^2} (\partial_\tau \phi)^2 + K_y a \cos^2 \phi
 \Bigr\},
\label{SGunif}
\end{equation}
where $N=N_A L/a$ is the total number of spins in the 
sample. Note that $\phi$ describes the azimuthal 
angle of a spin and  is a compact 
variable ($\phi+2 \pi$ is identified with $\phi$).
The tunneling amplitude between the anistropy minima at
$\phi=\pm \pi/2$ is then given by 
\begin{equation}
\langle \phi={\pi\over 2}|e^{-\beta {\cal H}}|\phi=
-{\pi\over 2}\rangle
=\int_{\phi(0)=-{\pi\over 2}}^{\phi(\beta)={\pi\over 2}}
 {\cal D} \phi\; e^{-\bar{\cal S}_{SG}[\phi]}.
\label{particletrans}
\end{equation}
The dominant contributions to the transition amplitude 
are the extrema of the action
which satisfy $\delta \bar{\cal S}_{\rm SG}=0$ or 
\begin{equation}
(J/c^2) \partial_\tau^2  \phi +  K_y \sin\phi \cos\phi=0\, .
\label{particleEL}
\end{equation}
Note that this equation is formally equivalent to 
(\ref{blochEL}). Similarly, as
a consequence of the symmetry of the action 
(\ref{SG})  under $\pi$-rotations around the 
hard-axis, i.e. $\phi\to\phi+\pi$, 
the two  anisotropy minima can be connected by 
two different paths.
These ``instanton" and ``antiinstanton" trajectories 
are given by $\phi_\pm(\tau)=-\pi/2 \pm 2 
\arctan e^{\omega_I(\tau-\tau_0)}$
and describe a transition from 
$\phi=-\pi/2$ to $\pi/2$ in clockwise 
($\phi_+$) or anticlockwise direction ($\phi_-$).
The transition occurs at $\tau_0$ 
within a finite imaginary 
time interval characterized 
by the ``instanton frequency" 
$\omega_I=c/\delta=(2a/S)\sqrt{K_yK_z}$.
Inserting $\phi_\pm$ into the 
action (\ref{SGunif}), we recognize
that the topological term gives rise to a phase 
which differs in sign for  instantons 
or antiinstantons
\begin{equation}
{\cal S}_{SG} [\phi_\pm]= \pm i \pi  N S + {\cal S}_0 ,   
\label{partinst}
\end{equation}
with the tunneling exponent\cite{lambda} 
${\cal S}_0 = 2 N S \sqrt{K_y/K_z}$.

The effect of the topological 
phase may now be seen as follows.
Adding the contributions of one single instanton
and antiinstanton to the action, we obtain 
for (\ref{particletrans})
\begin{eqnarray}
\langle \phi={\pi\over 2}|e^{-\beta {\cal H}}|
\phi&=&-{\pi\over2}\rangle
\propto \sum_{\sigma=\pm 1} e^{-{\cal S}_{SG}[\phi_\sigma]} \nonumber\\
&=&2 \cos(\pi NS) e^{-{\cal S}_0}\,.
\end{eqnarray}
Thus the transition amplitude vanishes
for half-odd integer $NS$ since 
tunneling paths of opposite winding
(or  ``chirality") interfere destructively 
with each other. A calculation within the
``dilute instanton gas" approximation \cite{raj}
reveals that this interference persists  
to all orders in the instanton contributions. 
Identifying $\mu=2N Ja/c^2=N S^2/ 2K_z a$,
$\kappa=N K_y a$, $\alpha=-NS$, $d=\pi$
we obtain (including the contributions
of Gaussian fluctuations around the instanton) 
\begin{equation}
\langle{\pi\over 2}|e^{-\beta {\cal H}}| 
-{\pi\over 2}\rangle
\propto e^{-\beta{\omega_I\over 2}}
\sinh \left( \beta {\Delta \over 2} \cos(N S \pi) \right).
\label{particleresult}
\end{equation}
with $\Delta=16 \sqrt{{N\over \pi S}} a 
\left(K_y^3  K_z\right)^{1/4} e^{-{\cal S}_0}$. 
Taking the limit $\beta \to \infty$ in Eq. 
(\ref{particleresult}) we conclude that
the ground state energy  of the 
individual potential wells $\omega_I/2$ is 
split into two levels separated by $\Delta$ 
provided the total spin $NS$
of the particle is  integer. 
For  $NS$ half-odd integer no splitting
occurs. Thus for arbitrary spin $S$
the  splitting $\Delta E$ between the two  states 
of lowest energy is given by 
\begin{equation}
\Delta E=|\cos(\pi NS)| \Delta.
\label{particlesplit}
\end{equation}
Thus, in the uniform limit our theory thus reproduces 
the spin-parity effect of Ref. \onlinecite{LDG}; 
moreover, in the  limit 
$K_y\ll K_z$, the tunneling exponent agrees with 
Refs. \onlinecite{korenblit,chudi} and $\Delta$  with 
Ref. \onlinecite{schilling}.

\section{Bloch Walls and Spin Waves}
\label{dissipation}

In this section we discuss the interaction between 
a  Bloch wall and its surrounding spin waves. 
We consider here a sample with $N_A$ (or $S$) 
sufficiently  large such that spin waves 
are just a small perturbation of the Bloch wall. 
For a quantitative description of this 
interaction we use a systematic approach 
with the  ratio of  wall velocity to 
spin wave velocity $\dot X/c$ as a small parameter. 
This is justified since 
typically $\dot X\ll c\sim 10^4 {\rm cm/s}$.
We construct then an ab initio theory for the 
soliton dissipation by integrating out the spin waves. 
Finally, by deriving the  spectral function 
of the damping kernel we can make contact
with the phenomenological Caldeira-Leggett 
formalism of dissipation.\cite{caldleg}
A brief account of the following results has been given in 
Refs.\onlinecite{BLii,BL}.

We consider elongated samples (cf. Fig. \ref{sample})
of sufficiently small cross sectional area
such that the transverse spin waves \cite{winter,janak}
around the Bloch wall are frozen  out. \cite{gap}
This condition typically requires $N_A$ to be less than $10^3$, and
thus can  easily be reconciled with above condition that $N_A\gg1$.
Motivated by  materials such as YIG which are favorable for MQP 
we consider the limit of large  hard-axis anisotropy. This allows
us to build upon the results of the last section 
and we can treat the interaction between 
Bloch wall and spin waves within the  
sine-Gordon model. \cite{footnote:general}

For notational simplicity
we restrict ourselves for the moment
to one of the Bloch walls (\ref{phiqc}),
$\phi_0(x) = \phi_{Q=1,C=1}(x)$. 
First we recall that $\phi_0(x-X)$ is, for arbitrary 
$X$, a static solution of $\delta{\cal S}_{SG}=0$. 
We now consider field configurations describing 
a Bloch wall at a position $X$ surrounded 
by arbitrary spin waves $\varphi$
\begin{equation}
\phi(x,\tau)= \phi_0(x-X) + \varphi(x-X,\tau),
\label{expand}
\end{equation}
and {\it elevate} $X(\tau)$ to a dynamical variable. 
However, Eq. (\ref{expand}) contains now a redundant
description of a rigid translation of the soliton:
A translation is either described by $X$ or by the 
``zero-mode" (Goldstone mode) 
$\varphi_0(x,\tau)\propto \phi_0'(x)$
\cite{dashen,gervais,sakita,raj}. 
To avoid double counting,
we thus have to impose the constraint that the 
spin wave modes be orthogonal  to the zero mode
\begin{equation}
\int dx\; \phi_0'(x) \varphi(x,\tau) =0,
\label{constraint}
\end{equation}
for all imaginary times $\tau$. We incorporate 
this constraint into the path integral by 
means of the Faddeev-Popov technique\cite{gervais,sakita} 
which we now briefly sketch. It is based on the identity 
\begin{equation}
\int {\cal D} X \delta(Q[X])  {\rm det} 
{\delta Q\over \delta X} =1,
\label{FP}
\end{equation}
with the judiciously \cite{sakita,zeromode} chosen functional
\begin{equation}
Q[X]=\int d x \; \phi_0'(x-X) \phi(x,\tau).
\label{Q}
\end{equation}
Inserting (\ref{expand}) into (\ref{Q}) 
we recognize that the $\delta$-function enforces 
the constraint (\ref{constraint}) as desired. 

For configurations which contain
one soliton,  we 
thus can rewrite the transition 
amplitude (\ref{SGpath}) as follows
\begin{equation}
\langle\{{\bf \Omega}_b\} |
e^{-\beta {\cal H}} | \{{\bf \Omega}_a\}\rangle = 
\int\!\!{\cal D} X \, {\cal D} \phi \;\delta(Q)
\;{\rm det} {\textstyle{\delta Q \over \delta X}}\;
e^{-{\cal S}_{SG}[\phi] }\, ,
\label{zba}
\end{equation}
where the action is given by (\ref{SG}).
We now perform a systematic expansion up to 
$2^{\rm nd}$ order in both spinwaves $\varphi$ and 
$\dot X/c$. [Note that $\dot X/c< 1.5 \cdot10^{-2}$ 
for YIG as discussed in Sec. \ref{expts} and 
$\varphi \propto 1/\sqrt{N_A}$ as we shall see below].
After insertion of (\ref{expand}) into (\ref{zba})
and expansion to second order in the 
spin waves $\varphi$ and second order in 
the Bloch wall velocity $\dot X/c$, the transition 
amplitude takes the form
\begin{equation}
\langle\{{\bf \Omega}_b\} |
e^{-\beta {\cal H}} | \{{\bf \Omega}_a\}\rangle= \int\!{\cal D} 
X\; e^{-S_X[X]}\; F[X],
\label{zf}
\end{equation}
where
\begin{equation}
{\cal S}_X = \int d\tau \Bigl\{ -i\alpha \dot X + {M\over 2}
\dot X^2 \Bigr\}
\label{sx}
\end{equation} 
is the action of a free Bloch wall, and where 
\begin{equation}
F[X] = \int \! {\cal D} \varphi \; 
\delta\!\biggl(\textstyle{\int} \phi_0' \varphi\biggr)
{\rm det} {\delta Q \over \delta X} 
e^{-N_A \{\varphi \cdot [{\cal G} 
+ {\cal K}]\varphi + {\cal J}\cdot \varphi \}}
\label{F}
\end{equation}
describes the interaction between the Bloch wall
and the spin waves. Here we have introduced the 
scalar product $a\cdot b=\int dx \; d\tau \; a^* b$ 
and the integral in the $\delta$-function is 
understood as an integral over $x$.

We now discuss the various terms that 
have been introduced in (\ref{zf})-(\ref{F}).
The first term in this action has 
the form of a gauge potential 
\begin{equation}
\alpha=\pi S N_A/a.
\end{equation}
It originates from the topological term in (\ref{SG})
and from the relation 
$\int  dx \; \partial_\tau \; \phi_0(x-X) = -\pi \dot X$
since each soliton flips the spins by 
$\pi=\int dx \phi_0'$. 

The second term in (\ref{sx}) is the kinetic 
energy of the Bloch wall and the  mass is given by 
\begin{equation}
M={E_0\over c^2} = {N_A S^2 \over a^2} 
\sqrt{{K_y\over J}} {1\over K_z}.
\label{mass}
\end{equation}
This value coincides with the D\"oring mass \cite{sloncz}
$M=N_A a^2/(2\pi \gamma^2 \delta)$ with $\gamma =g \mu_B/\hbar$
if the hard-axis anisotropy is of purely demagnetizing origin,
$K_z=2 \pi M_0^2 a^2$ with $M_0=g \mu_B S/a^3$.
We thus have given a {\it microscopic derivation of 
the D\"oring wall mass}.

In (\ref{sx})  we have dropped a term $\beta E_0$ with 
$E_0$ the Bloch wall energy (\ref{wallenergy})
since the Bloch wall already exists in the 
sample and is not created thermally. 
The thermal creation of Bloch wall pairs 
in the absence of an external field is 
negligibly small for temperatures in the Kelvin range 
even for samples as small as $50 \AA \times 50\AA$.
Only at higher temperatures and in the presence of external 
fields, thermal creation of Bloch wall pairs becomes 
appreciable.\cite{prb2} 
 
The functional $F$ (\ref{F}) describes the coupling 
between spin waves and the Bloch wall. The operator\cite{prb1}
\begin{equation}
{\cal G} = -J\partial_x^2 -\kappa \partial_\tau^2 + 
K_y [1 - 2 {\rm sech}^2 ({x \over \delta})]
\label{G}
\end{equation}
with $\kappa = J/c^2$
describes the spin wave spectrum around 
a static Bloch wall. 
The remaining operators
are responsible for the dynamic coupling 
between spin waves and domain wall,
\begin{eqnarray}
{\cal K} &=& 2 \kappa \dot X \partial_x \, \partial_\tau 
- \kappa \dot X^2 \partial_x^2, \qquad
{\cal J} = - 2 \kappa \dot X^2 \phi_0''.
\end{eqnarray}
Due to the constraint (\ref{constraint}) the exponential 
in $F$, Eq. (\ref{F}), does not contain 
a term linear in the velocity 
$\dot X$ and in the spin waves $\varphi$. 
It has been pointed out \cite{stamp,scb} 
that this is an important difference to the 
standard Caldeira-Leggett model.\cite{caldleg}
However, despite this non-linear coupling 
we shall see shortly that at low temperatures the  
dissipation due to spin waves  can 
--- if this should be desirable --- 
perfectly well be modeled by a Caldeira-Leggett approach 
(such cases have actually been discussed in App. I of 
Ref. \onlinecite{caldleg}), although the precise 
form of the relevant spectral function can 
only be obtained from a microscopic calculation 
as presented here. 

After the evaluation of (\ref{zba}) which is 
sketched in  Appendix \ref{appendixb}, 
and collecting Eqs. (\ref{sx}), (\ref{Fdet}),
(\ref{trexpand}), and  (\ref{damp2}) we can 
express the transition amplitude (\ref{zf}) as
\begin{equation}
\langle \{ {\bf \Omega}_b \} |e^{-\beta {\cal H}}| \{{\bf \Omega}_a\}
 \rangle  = \int\! {\cal D} X \; e^{-S_{\rm eff}[X]},
\end{equation}
with the effective action for the soliton position, 
\begin{eqnarray}
S_{\rm eff}[&X&]=\int_0^\beta \! d\tau \Bigl\{ -i \alpha \dot X + 
{M_{\rm eff}\over 2} \dot X^2 \Bigr\}+\nonumber\\ 
&+&{1\over 2} \int_0^\beta \! d\tau
\int_0^\tau\!d\sigma K(\tau-\sigma) [X(\tau)- X(\sigma)]^2.
\label{seff}
\end{eqnarray}
The damping kernel has been evaluated for arbitrary 
temperatures in Appendix \ref{appendixb}. Here we  
restrict ourselves to low temperatures, $\beta \to \infty$, 
where the damping kernel (\ref{K}) takes the following form 
\begin{equation}
K(\tau)=-2 \sum_k k^2 \omega_k^2 e^{-2\omega_k|\tau|}\,.
\label{Ksum}
\end{equation}
This can also be cast into standard Caldeira-Leggett 
notation \cite{caldleg,weiss}
\begin{equation}
K(\tau) = {1\over \pi} \int_0^\infty \!\! d\omega \; J(\omega) 
D_{\omega}(\tau).
\label{Klowtemp}
\end{equation}
Here $D_{\omega}(\tau) = e^{-\omega |\tau|}$ 
is the $T\to 0$ limit of (\ref{D}) and  
the spectral function is given by \cite{specfunc}
\begin{equation}
J(\omega)={\omega\over \omega_0 \delta^2} 
\; \Theta(\omega-\omega_0)  \sqrt{\omega^2 - \omega^2_0}\, ,
\label{jomega}
\end{equation}
which vanishes for $\omega<\omega_0\equiv 2 c/\delta
=(4a/S)\sqrt{K_y K_z}$,
the anisotropy gap of the spin waves.
[For material values as in YIG, this gap corresponds
to a temperature of $T_g=0.2 K$. Other materials 
have in general larger anisotropies and thus higher $T_g$.] 
In deriving (\ref{jomega}) from (\ref{Ksum}) we have 
used the renormalization (\ref{sumrenorm}). It is only after this 
renormalization that the  memory kernel $K$  becomes 
positive definite (as is needed for convergence).

If now the dynamics of $X$ is slow compared to the 
time variation of the damping kernel, i.e. if the instanton
frequency $\omega_I$ (to 
be evaluated below)  is much smaller than $\omega_0=2c/\delta$,
and if the temperature is  small such that $\beta \gg \omega_I^{-1}$
then we may expand $X(\tau) - X(\sigma) \approx (\tau -\sigma)
\dot X(\sigma)$,
and the damping kernel reduces to a pure mass renormalization. Note 
that this mass renormalization is a ${\cal O}((N_A)^0)$
correction of the wall mass $M\propto N_A$. Since all these 
conditions will be
satisfied for the tunneling situations considered below, we see
that we end up with a deceptively simple
effective description of the Bloch wall dynamics, given
by the first two terms in (\ref{seff}).

\section{Interference effects due to the  Berry  phase}
\label{interference}

In the last section we derived an effective action 
for the dynamics of the Bloch wall position. We showed 
that damping  due to spin waves leads to a gap in the 
spectral function and thus leads to a mere  renormalization 
of the wall mass at low temperatures. 
More importantly, we  have identified 
a topological term in the action which has its origin in 
the Berry phase term (\ref{swzintro}) 
of the original spin action. 

Here we shall generalize these 
considerations to  solitons $\phi_{QC}$  of 
arbitrary chirality $C$ and charge $Q$ which 
are all energetically degenerate. 
We show that the interference between 
states of different chirality gives rise 
to remarkable  effects such as the 
halving of the Brillouin zone and the alternation of chirality in
reciprocal space for half-integral spin. 

These effects originate in the 
fact that for arbitrary solitons, the topological term 
(\ref{swzintro}) 
\begin{equation}
i{\alpha \over \pi} \int dx \dot \phi_{QC}(x-X)
=-i \alpha C \dot X,
\label{alphac}
\end{equation}
depends on the soliton chirality $C$ (but not on its charge $Q$)
($\alpha= \pi S N_A/a$). This chirality dependence 
can intuitively be understood as follows. As the soliton sweeps 
across a given spin, the spin is rotated by 
an angle $\pm\pi$ (after the wall is sufficiently far away),
the rotation sense being
uniquely determined by the chirality and the 
direction of motion of the Bloch wall.

At low temperatures and for solitons of 
arbitrary chirality the effective action
(\ref{seff}) of a soliton in 
an external potential $V(X)$ thus takes the following form 
\begin{equation}
{\cal S}[X,C] =\int d\tau \Bigl\{
-i\alpha C \dot X + {M\over2} \dot X^2 + V(X)\Bigr\}.
\label{xaction}
\end{equation}
Here we have used the mass (\ref{mass}) rather than the 
dressed mass $M_{\rm eff}$ since the 
mass renormalizations at low temperatures are 
small ${\cal O}((N_A)^0)$ and the value (\ref{mass})
thus represents a good approximation for the 
experimentally observed wall mass. 
In addition we have introduced a periodic potential 
$V(X)$ of period $d$, and we make the natural
assumption that $d$ is some integer multiple of the
lattice constant $a$.
For definiteness we assume 
\begin{equation}
V(X) =  V_0[1- \cos \left({2\pi X \over d}\right)]\, ,
\label{V}
\end{equation}
which has  amplitude $2V_0$. \cite{pinning}
Such a potential can have its origin \cite{BL,BLii} 
in the discrete nature of the crystal 
lattice itself, or for Bloch walls
it can arise from a magnetic superlattice 
of layers with different
anisotropies.

The action (\ref{xaction}) corresponds to the 
Hamiltonian 
\begin{equation}
{\cal H}={1\over 2M} (p-\alpha \sigma_z)^2 + V(X),
\label{heff}
\end{equation}
where $p=-i\partial/\partial X$ 
is the Bloch wall momentum, 
$\sigma_z$ is the Pauli-matrix of the 
``pseudospin" characterizing the 
chirality $C=\pm$ of the Bloch wall. 
Obviously, this Hamiltonian conserves the chirality.
For mathematical convenience,
we choose periodic boundary conditions in the following.
However, all our results are finite in the 
thermodynamic limit and none of our 
conclusions depend on this
choice of boundary conditions. \cite{footnote2walls}

{}From both  (\ref{xaction})
and (\ref{heff}) it is evident that 
the topological phase plays the role 
of a gauge potential  
whose effect on the wall dynamics
shall be discussed next. 
We note that such spin-dependent gauge potentials 
are not uncommon in problems involving Berry phases. \cite{Lgold}

\subsection{An illustrative example}
\label{interference1}

Before giving a rigorous discussion of the
dispersion relation, we  give an  argument
to illustrate the interplay  between 
the topological phase and soliton propagation.

Consider the transition amplitude 
for the propagation of the Bloch wall 
between nearest neighbors which is given by 
\begin{eqnarray}
\langle 0 | e^{-\beta {\cal H}}| d \rangle
&=& \sum_C \int_0^d {\cal D} X 
e^{i\alpha C \int \! d\tau \;\dot X}    
e^{-{\cal S}_0}, \nonumber \\
&=& 2\cos (\alpha d) \int_0^d {\cal D} X
e^{-{\cal S}_0},
\label{hop}
\end{eqnarray}
where ${\cal S}_0 [X]= 
\int d\tau \{ (M/2) \dot X^2 + V(X) \}$. 
For half-integer $\sigma\equiv N_A S d/a$, 
 we thus arrive at a most important conclusion: 
{\it Nearest neighbor hopping of the soliton 
is suppressed if both chiralities 
contribute equally to the transition amplitude.} 
However, if the soliton is in a state of 
{\it definite chirality}, only one path 
contributes to the transition 
amplitude (\ref{hop})  and {\it nearest 
neighbor hopping is allowed.}
No such interference occurs for integer $\sigma$.
Note that this interference effect is entirely
due to the topological term in  (\ref{xaction}) 
which in turn is a consequence of the topological 
term in the  sine-Gordon action (\ref{SG}). 

We  now  investigate how this interference 
affects the dispersion of solitons
and in this way can become observable.

\subsection{Dispersion in the nearly free limit}
\label{interference2}

In this section we discuss the dynamics of a
soliton in an arbitrarily weak
periodic potential $V(X)$. The Hamiltonian is 
given by (\ref{heff}) with
$V_0 \to 0$. Despite being simple
this case already captures most of the characteristic
features of the tight binding limit which shall be 
discussed below.  For simplicity  we assume 
that the period of the potential is given by $d=a$.

Using periodic boundary conditions, the eigenstates 
of (\ref{heff}) are  simply plane waves 
$e^{ikX}$, with $k=2\pi n/L$, $L=N d$, and the spectrum 
consists of two 
parabolas (corresponding to the two soliton-chiralities)
\begin{equation}
E(k,C=\pm 1)={1\over 2M} (k \mp \alpha)^2,
\label{efree}
\end{equation}
periodically extended by the reciprocal lattice 
vector $G=2\pi /a$. 
[Note that the requirement of gauge invariance 
alone produces such a periodic extension even
in complete absence of a periodic potential.
The gauges (\ref{gauge}) lead to 
Hamiltonians (\ref{heff}) with $\alpha \to (2n+1)\alpha$ with 
$n=0,\pm1,\dots$. The gauge invariant 
dispersion is therefore the  periodic extension 
of (\ref{efree}) by a vector $2\alpha$.]

The Berry phase thus leads to remarkable
spin parity effects in the dispersion:
For half-odd integer spin $N_A S$, 
we have $\alpha=G/4$ (mod $G$) and the parabolas
are separated by half the reciprocal lattice vector, $G/2$.
Thus,
the Brillouin zone is halved
and two subsequent parabolas belong to opposite chiralities
as illustrated in Fig.\ref{freedisp}.
The observability of this is discussed in Sec. \ref{interference4}.

For integer spin, however, the dispersion is 
analogous to that of a free particle of mass $M$ and 
the Berry phase is inoperative since it merely shifts 
the dispersion by a reciprocal lattice vector.

Note that since $\alpha \propto 1/a$ is 
independent of the sample 
length, the 
result is unchanged if we pass to the thermodynamic
limit. Therefore (\ref{efree}) is independent of the 
boundary conditions. \cite{footnotepc}

\subsection{Tight binding limit}
\label{interference3}

We now turn to a discussion of the system
in the tight binding limit where $V(X)$
is no longer small. In the absence of 
tunneling  there exists
a large number of degenerate ground states corresponding 
to the soliton  trapped at one particular pinning site.  
If the pinning potential is not too large,
the soliton can tunnel between the sites,
and these ground states split into the 
(lowest) band $E(k,C)$ with
\begin{equation}
{\cal H} |k,C\rangle = E(k,C) |k,C\rangle.
\label{eigenvalue}
\end{equation}
Since ${\cal H}$ in (\ref{heff}) is invariant under 
translations by the potential period $d$ and 
conserves the chirality pseudospin,
the eigenstates are products of Bloch states 
and chirality eigenstates 
$|k,C\rangle =|k\rangle\bigotimes |C\rangle$
where  ${\cal T} |k\rangle = e^{ikd} |k\rangle$, 
$\sigma_z |C\rangle = C |C\rangle$
with $k=2\pi n/Nd$, $n=0,1,...,N-1$, and $L=Nd$.
${\cal T}$ is the translation operator 
$\langle X|{\cal T} = \langle X+d|$.

For the evaluation of the band structure
in the tight-binding limit we now
develop a formalism which allows us
to keep carefully track of the topological phases 
within the instanton approach. 
To this end we start from the modified partition function
\cite{zinn}
\begin{equation}
{\cal Z}_l={\rm tr} \left\{ {\cal T}^l 
e^{-\beta {\cal H}}  \right\},
\label{zl}
\end{equation} 
where ${\rm tr}\{\cdot\}=\sum_{k,C} \langle k,C|\cdot|k,C\rangle$.
We use (\ref{zl}) rather than the usual partition 
function ${\cal Z}={\rm tr}\{e^{-\beta {\cal H}}\}$ 
for the following reason. As we have seen 
in the previous section, the Berry phase gives rise to 
a  shift of the dispersion with respect to $k$. However, 
the partition function ${\cal Z}$ is insensitive to 
such shifts (at least if there is no perturbation which mixes 
the chirality states) and thus represents 
an insufficient tool for the evaluation of the band structure.  

{}From (\ref{zl}) we can easily 
extract the dispersion by taking the Fourier transform 
\begin{equation}
\sum_{l=0}^{N-1} e^{-ikld} {\cal Z}_l = N \sum_{C=\pm} e^{-\beta E(k,C)},
\label{modz2}
\end{equation}
where we used
(\ref{eigenvalue}) and the definition of ${\cal T}$.
In addition, we  have restricted ourselves 
to the lowest band since we are interested in the 
low temperature limit. 
Note also that the lhs of (\ref{modz2})
does not contain higher winding number contributions 
since we are not interested in finite size effects 
arising from the sample topology. 
We now evaluate  ${\cal Z}_l$ and get in a first step, 
\begin{eqnarray}
{\cal Z}_l
&=&\sum_C  \int_0^L dX \; 
\langle X+ld,C|e^{-\beta{\cal H}} |X,C\rangle \nonumber \\
&=&\sum_{C,k}\sum_{m=0}^{N-1}  \int_{md-d/2}^{md+d/2}\!\!\! dX \; 
|\langle X,C|k\rangle|^2
e^{-\beta E(k,C)+ ikld} \, ,
\end{eqnarray}
where we used periodic boundary conditions and inserted a complete
set of Bloch states. Next, in the tight binding limit
the main contributions to the integral are coming from
the vicinity of the potential minima, $X\simeq md$, 
and the Bloch functions can be replaced by
their harmonic approximations, i.e.
$|\langle k|md\rangle|^2\approx|\psi_{\rm h}(0)|^2/N=a_0/N$.
Here, $\psi_{\rm h}$ is the ground state in the harmonic approximation 
of the potential well and $a_0=\sqrt{M\omega/\pi}$ 
its normalization (squared). Thus, we find
\begin{equation}
{\cal Z}_l
\approx {1\over a_0} \sum_C \sum_{m=0}^{N-1} 
\langle md+ld,C | e^{-\beta {\cal H}} |md,C\rangle\, .
\label{xcont}
\end{equation}
Using a  path integral representation 
for (\ref{xcont}) and employing the periodicity of ${\cal H}$
we obtain with (\ref{modz2}) 
\begin{eqnarray}
\sum_C e^{-\beta E(k,C)} = {1\over a_0} \!\sum_{C,l=0}^{N-1} 
e^{-ikld} \int_{X(0)=0}^{X(\beta)=ld}\!\!\!\!\!{\cal D}X\;
e^{-{\cal S}[X,C]}.
\label{zpath}
\end{eqnarray}
The path integral on the rhs of (\ref{zpath}) is  dominated 
by instantons between the 
potential minima.  These instantons obey the 
Euler-Lagrange equation 
$\delta {\cal S}/\delta X = -M \ddot X + V'(X)=0$. 
For instance, a transition from $X=0$ to $X=d$ ($X=-d$) is 
mediated by the (anti-) instanton 
\begin{equation}
X^\pm = \pm {2d\over \pi} \arctan\; e^{\omega (\tau-\tau_0)},
\end{equation}
centered at the arbitary imaginary time $\tau_0$. 
The instanton frequency 
$\omega={2\pi \over d} \sqrt{V_0/M}$
equals the harmonic oscillation frequency in the potential well. 
The instanton action is given by 
\begin{equation}
{\cal S}_{\pm}={\cal S}[X^\pm,C]= {\cal S}_0 \mp i\alpha d C,
\label{sxpm}
\end{equation}
where ${\cal S}_0={4\over \pi} d \sqrt{M V_0}=8 (V_0 / \omega)$.
The unusual second term in (\ref{sxpm}) 
is purely imaginary and is a direct 
consequence of the 
gauge potential in (\ref{seff}) or 
(\ref{heff}) and distinguishes 
between instantons and antiinstantons.  
Note that this term does not 
break time reversal invariance as the 
partition function contains contributions 
of both chirality states $C=\pm 1$.

The path integral in (\ref{zpath}) can be expressed as the 
sum over all distinct sequences of
$n_+$ instantons and $n_-=n_+-l$ antiinstantons
which connect the initial state $X=0$ with the 
final state $X(\beta)=ld$. Within this 
``dilute instanton gas approximation" \cite{raj} we obtain 
\begin{eqnarray}
\int_{X(0)=0}^{X(\beta)=ld} &{\cal D}X&\;
e^{-{\cal S}[X,C]}=
a_0 e^{-{\beta \omega\over 2}}
\sum_{n_+=0,n_-=0}^\infty\delta_{n_+,n_- +l} \,\, \times \nonumber \\
&\times&{(JK\beta e^{-{\cal S}_+})^{n_+}\over n_+!}
{(JK\beta e^{-{\cal S}_-})^{n_-}\over n_-!}, 
\label{instsum} 
\end{eqnarray}
where $J=\sqrt{{\cal S}_0/2\pi M}$, $K=2\omega \sqrt{M}$ 
arise from the integration over the zero modes
and the Gaussian fluctuations around an instanton,
respectively. 
Inserting (\ref{instsum}) into (\ref{zpath}), using 
(\ref{sxpm}) and performing the sums we obtain  
\begin{equation}
\sum_C e^{-\beta E(k,C)} =\sum_C e^{-\beta 
({\omega\over 2}+\epsilon(k,C))}\, ,
\label{essence}
\end{equation}
where
\begin{equation}
\epsilon(k,C)= - 2 J K e^{-{\cal S}_0}
\cos\Bigl((k+\alpha C) d\Bigr).
\label{band1}
\end{equation}
The ground state as a function of $k$ is given by 
\begin{equation}
E(k) = - \lim_{\beta \to \infty}{1 \over \beta} \ln  
\sum_C e^{-\beta E(k,C)}.
\label{ground}
\end{equation}
Similarly to the  nearly free limit discussed in
the previous subsection, this dispersion is 
fundamentally different for 
$\sigma\equiv\alpha d/\pi= N_A S d/a$ 
integer or half-odd integer. \cite{BLii} 

Inserting (\ref{essence}) into (\ref{ground}) we obtain
for integer $\sigma$ the following dispersion
\begin{equation}
E(k) = - {\Delta \over 2} \cos(kd),
\label{ekint}
\end{equation}
which is of standard tight-binding type. In 
(\ref{ekint}) we dropped the constant $\omega/2$.
The bandwidth is given by
\begin{equation}
\Delta=8 \omega 
\sqrt{{\cal S}_0 \over 2\pi} e^{-{\cal S}_0}.
\label{bandwidth}
\end{equation}

In contrast, for $\sigma$ half-integer 
we obtain 
\begin{equation}
E(k) = - {\Delta\over 2} | \sin(kd) |.
\label{ekhalf}
\end{equation}
In (\ref{ekint}) and  (\ref{ekhalf}) 
we have suppressed sign changes which 
correspond to a global shift of $k$ by $\pi/d$. 
Such a global sign cannot be measured since the 
absolute value of $k$ is experimentally not detectable. 

The dispersion (\ref{ekhalf})  has now cusps and 
{\it the bandwidth and the Brillouin zone  
are  halved} \cite{footnote:halving}, as shown in 
Fig.\ref{tightbinding}. Moreover, we draw from (\ref{essence})
and (\ref{ground}) the important conclusion
that {\it states whose wavevector 
differs by $\pi$  have  opposite chirality}, 
cf. Fig.\ref{tightbinding}.

Note that this period halving in reciprocal space 
is a consequence of the fact that ${\cal Z}_l = \sum_C \int_0^{ld} 
{\cal D} X e^{-{\cal S}[X,C]}=0$ for $l$ odd (cf. (\ref{hop})). 
However,  one must not conclude from this 
fact that nearest 
neighbor hopping is always suppressed: At fixed 
$k\neq 0,\pm{\pi\over d}$, the ground state condition
(\ref{ground}) selects a branch 
of the dispersion with {\it definite} chirality, 
a dispersion that arises from nearest 
neighbor hopping. Only at the cusps in (\ref{ekhalf}) 
hopping is suppressed. 

This band-halving can also be understood in a more 
intuitive way: For half-integer $\sigma$, a soliton
acquires a Berry phase $-iC$ for forward ($iC$ for backward) 
hopping. In the ground state this phase gets compensated
by the Bloch phase thus creating two band
minima at $kd=\pm \pi/2$ which have opposite chirality.

Finally we give a more explicit formal argument for 
the chirality correlation. We find the explicit
form of the eigenvalue $E(k,C)$ by
repeating the steps leading to Eq. (\ref{essence}),
but instead of ${\cal Z}_l$ we use 
${\cal Z}_l^C={\rm tr}\{ |C\rangle\langle C| 
{\cal T}^l e^{-\beta {\cal H}} \}$, 
which projects onto a state of definite chirality $C$.
Thus, we find that $E(k,C)$ is given by $\epsilon(k,C)$ in
Eq. (\ref{band1}).
By comparing the ground state energy (\ref{ekhalf}) 
with $E(k,C)$ we see that $k$-intervals with positive 
(negative) $\sin kd$ 
belong to negative (positive) chirality $C$.
This result is derived in the north pole parametrization.
If, instead, we use the south pole parametrization, 
then the gauge potential in (\ref{xaction}) 
changes sign and again we find that the chirality alternates, but now
with the opposite assignment between chirality and given $k$-interval. 
The physical consequence---alternating chirality with changing $k$---
is the same in the two gauges, since, again, the absolute $k$ value 
cannot be observed. 
 
\subsection{Discussion and analogies to 
other physical systems} 
\label{interference4}

In the last two sections we have seen that 
the dispersion is strongly affected by the  
parity  of $\sigma=N_A S d/a$. For 
$\sigma$ integer, the dispersion equals that of 
a particle  in a periodic potential while
for $\sigma$ half integer a halving of the Brillouin zone
occurs with alternating chiralities.
In the latter case the dispersion consists of mutually intersecting 
parabolas or tight binding bands. How can we observe such a dispersion? 

Let us for definiteness focus on the nearly free limit
with a dispersion as shown in Fig. \ref{freedisp}.
Suppose the chirality has been measured to be $C=1$ and 
the system is in its ground state, i.e. in the 
minimum of a $C=1$ parabola. 
If we now drive the system out of its
energy minimum, e.g., by applying an external field 
along the easy-axis (see below), the Bloch wall 
will follow the $C=1$ parabola.
The Bloch wall will remain on this parabola 
even beyond the crossing point, {\it provided} that the  
chirality $C$ is a  
conserved quantity. In this sense, the two 
parabolas for $C=1$ and $C=-1$ behave like two 
``Riemann sheets" of the energy which 
are completely disconnected, and  their intersection
has no observable consequences if there is no mixing i.e. tunneling 
between the chiralities of the Bloch wall. 

Nevertheless, the dispersion of Fig. \ref{freedisp} (thin line)
is a precursor of a striking physical 
effect: As soon as there is 
tunneling between the chiralities, the different ``sheets"
get connected and for half 
integer spins a gap develops at the crossing 
points of parabolas belonging to $C=\pm 1$. 
At the same time the halving of the Brillouin zone becomes 
observable.
Formally this can be described as follows.
In the presence of tunneling between the two 
wall chiralities (see Sec. \ref{expts}) 
the Hamiltonian (\ref{heff}) acquires an additional term 
$\epsilon \sigma_x$
\begin{equation}
{\cal H} = {1\over 2M} (p - \alpha  \sigma_z)^2 
+ V(X) + \epsilon \sigma_x,
\label{hsigmax}
\end{equation}
such that the chirality $C$ (i.e. $\sigma_z$) is no longer 
a conserved 
quantity. We are interested in the limit of small chirality 
tunneling and therefore  $\epsilon$ will be much smaller than 
the bandwidth $\Delta$ (estimates for $\epsilon$ will be given
in Sec. \ref{expts}). For $\sigma$ half-odd  integer, 
the degeneracy at the points $k_n=n\pi/a$ 
is lifted and the dispersion splits into two bands
which for $|k-k_n|\ll \pi/a $ are given by 
\begin{equation}
E_{\pm}(k) = {1\over 2M} \left[(k - k_n)^2 + \alpha^2 \pm 
\sqrt{4 \alpha^2 (k - k_n)^2 + \tilde\epsilon^2}\right],
\label{epm}
\end{equation}
with $\tilde \epsilon =  2M \epsilon$ and 
where, for simplicity, we have stated the result 
in the nearly free limit. In this 
and the tight binding limit the two bands 
are  separated  by a gap $2\epsilon$ at $k=k_n$, as shown in 
Figs. \ref{freedisp},\ref{tightbinding}.

Solving for the corresponding eigenstates 
we recognize that the chirality continuously switches 
from $C=\pm 1$ to $C=\mp 1$ as we pass from one
band minimum to an adjacent one. 

We thus have established that the spectrum 
given in (\ref{efree}), (\ref{ekhalf}) is reached 
in the limit $\epsilon \to 0$.
Note that the experimental observation of the gap 
depends on the probability of Zener interband transitions
and thus on the time scale at which the  band structure
is probed. In the nearly free limit, 
the Zener probability \cite{zener} can be expressed 
as \cite{BL,BLii} $P\propto \exp \{- {\pi\over 2} {\epsilon^2 T
\over \hbar E_0}\}$, where $T=2\pi/\omega_B$ is the time to cross
the Brillouin zone, with $\omega_B=Fd/\hbar$ being the Bloch 
frequency and $F=2g\mu_B S N_A H/a$ the driving force due to
an external field $H$ (along, say, the easy  axis, see 
Sec. \ref{extfields}).
$E_0={\hbar^2\over 2M} (\pi/d)^2$ is the kinetic energy 
at the zone boundary. 
Thus, to optimize observability we must have  
$A={\pi\over 2} {\epsilon^2 T
\over \hbar E_0}\gg 1$, which is easy to achieve since typically
$T\sim 10^{-7}s$, giving $A\sim 100$ for YIG, if we choose 
$H\sim 10^{-3} {\rm Oe}$, $d=a$, and $\epsilon\sim
E_0/10\sim 10 {\rm mK} k_B$ (see Sec. \ref{expts}). 
The alternation of chirality could then be observed, for instance,
by (optical) dichroism techniques which would be sensitive to
the rotation sense of the magnetization within the Bloch wall.

We emphasize that these results are gauge independent.
If, instead, we had started from the south pole parametrization 
of the coherent states, we would have obtained the 
same dispersion  (\ref{efree}), (\ref{ekint}), (\ref{ekhalf}),
except for a global shift $k \to k+2\alpha$ which is unobservable.

A dispersion consisting of disjoint 
parabolas dictated by gauge invariance 
and the formation of gaps 
due to tunneling 
is quite a common phenomenon in condensed 
matter physics. 
Persistent currents in isolated metal rings \cite{persist}, 
the Josephson effect \cite{josephson}, 
and the tunneling of 
quasiparticles between edge states in the fractional 
quantum Hall regime \cite{thouless}
might serve as familiar examples.  

For further illustration let us briefly discuss
some relations between our spin effect and, say,
persistent currents. First, in
the spin system the dispersion 
remains unaltered in the thermodynamic limit, whereas 
persistent currents are 
a finite size effect, resulting from the 
discreteness of the energy levels. 
In addition,  we consider a simply connected 
sample topology while a persistent current 
relies on the ring geometry of the sample. 
In the spin system it is the $S^1$-topology of 
spin space  restricted to the easy-plane, 
not the topology of the sample which is 
responsible for the interference effect.

An electron of mass $m$ confined to a ring of
radius $\varrho$ which is threaded by the electromagnetic 
flux $\Phi$  is described by the Hamiltonian 
\begin{equation}
{\cal H} = {\hbar^2\over 2 m \varrho^2} 
(-i\partial_\theta -\Phi)^2\, , 
\end{equation}
where $\Phi$  is measured in units of
the flux quantum  $\Phi_0=hc/e$, and 
$\theta$ is the azimuthal angle. 
The eigenfunctions are $e^{in\theta}$ with eigenvalues
$E_n = (\hbar^2 / 2 m \varrho^2) (n-{\Phi})^2$, 
where $n=0,\pm 1,...$ The ground state energy $E_G$
as a function of flux is the envelope of the 
set of energy parabolas separated by $\Phi_0$. \cite{byersyang}
Thus, the persistent current $j=-(e/\hbar)
\partial E_G/\partial \Phi$ is a sawtooth curve 
with discontinuities at $|\Phi|=n/2$ where the parabolas
intersect. Suppose 
now that $\Phi=0$
and that the system is in its ground state with $n=0$. 
If the flux is increased  adiabatically, 
the system will stay on the $n=0$ parabola even 
for $\Phi > 1/2$  since the angular momentum is a 
conserved quantity. Thus the electron 
will not see the other parabolas and the spectrum
consists of disconnected ``sheets" of parabolas.
This behavior is analogous to that for the soliton
dispersion (\ref{efree}) for half-integer spin. 

However, if angular momentum is no longer conserved, 
e.g., due to the presence of a 
scattering potential, the parabolas will be
connected and a gap  develops 
at their crossing points. 
The scattering potential thus plays a role similar
to the $\epsilon \sigma_x$-term in (\ref{hsigmax})
caused by tunneling between the chiralities.

The mere existence of  interference effects in a 
metal ring can also be derived from the following 
argument. Assume that $\Phi=1/2$ and let 
us imagine having prepared two wave packets  
of opposite angular momentum, but otherwise identical. 
If we let these
wave packets dynamically evolve until they have
traveled half the circumference, one clockwise
the other anticlockwise, they will have picked up 
Aharonov-Bohm phases of opposite sign such that 
(for $\Phi=1/2$) destructive interference
occurs leading to a vanishing transition amplitude
between initial and final state. This behavior
is similar to the spin case described in (\ref{hop}),
where the clockwise 
and anticlockwise traveling wave packets correspond 
to the two chirality states of the soliton (note 
again that the real space topology of the ferromagnet 
is irrelevant). 

\section{experimental implications}
\label{expts}

In this section we give  numerical estimates for
the effects discussed in the previous sections.
For definiteness we concentrate on material parameters
for YIG.
Exchange \cite{tebble} and anisotropy \cite{dell} are given by 
$J=1.65\cdot 10^{-21} {\rm erg}\cdot {\rm cm}$, and
$K_y =9.61\cdot 10^{-11} {\rm erg}/{\rm cm}$, 
where a cell with lattice constant $a=6.2 \AA$
contains one $S=5/2$ spin implying a saturation 
magnetization \cite{dell} of $M_0=194 {\rm Oe}$
(i.e. $K_z=2\pi M_0^2a^2=9.1\cdot10^{-10} {\rm erg}/{\rm cm}$),
wall width  $\delta=\sqrt{J/K_y}=414\AA$, and spin wave 
velocity, Eq. (\ref{spinvel}),
$c=6 \cdot 10^4 {\rm cm/s}$.
The pinning potential strength can be related to an experimentally 
observed coercivity  by adding \cite{riehemann} a Zeeman term 
$-2 {\cal A} M_0 H_{\rm ext} X$ to the pinning potential $V(X)$, with 
${\cal A}=N_A a^2$ the cross sectional area of the sample.
Defining the coercivity $H_c$ as the field at which the barrier height 
vanishes, we obtain $V_0/{\cal A}=H_c M_0 {d/ \pi}$. 
Note that  the coercivity is proportional to 
the slope $V_0/d$ of the potential.  Looking at the 
WKB exponent (\ref{sxpm}), ${\cal S}_0=
{4\over \pi} d \sqrt{M V_0}$, 
we see that a low coercivity does not necessarily imply a high 
tunneling probability.  The crucial condition is 
a small potential width $d$.

We now assume a coercivity \cite{coercivity}
of $H_c=2 {\rm Oe}$ and $d=3a$. Note that the wall extends 
over 22 pinning sites. 
The instanton frequency then  becomes 
$\omega={2\pi \over d} \sqrt{V_0/M}=1.4\cdot10^{10} s^{-1}$,
and $|\dot X/c|= \omega_I d/\pi c\leq 1.5\cdot 10^{-2}$.
For a sample with cross section 
${\cal A}=10^4\AA$ we have $N_A=260$, and the wall contains
$N_A\delta/a=2\cdot 10^4$ spins.
The pinning potential
height takes the value $2V_0=330 {\rm mK}k_B$, and the  
bandwidth (\ref{bandwidth})
is $\Delta/\hbar \simeq  10^6 s^{-1}$,
which is of the order of the measured resonance frequency
in Ref. \onlinecite{awschalom}. 
The  D\"oring mass, Eq. (\ref{mass}), corresponding to this cross 
sectional area ${\cal A}$ takes the value
$M=1.24\cdot 10^{-22}{\rm g}=1.36\cdot 10^5
{\rm m_e}$, where ${\rm m_e}$ is the electron mass.
The crossover temperature between quantum tunneling 
and thermally activated behavior is $T_c=2V_0\hbar/{\cal S}_0 k_B
=\hbar\omega/4k_B
\simeq 28 {\rm mK}$, since $\hbar\omega/k_B=110 {\rm mK}$ for $d=3a$. 
Note that the bandwidth is extremely 
sensitive to the details of the pinning potential. For instance,
if $d=a$ (lattice pinning) but all other parameters
chosen as above, we obtain $\omega=2.5 \cdot 10^{10}s^{-1}$ 
(corresponding to $190 {\rm mK}$), and
$\Delta/\hbar \simeq 1.2 \cdot 10^{10} s^{-1}$ 
(since ${\cal S}_0/\hbar=2.3$),
or  $\Delta \simeq 0.8$ times the pinning 
potential height  $2V_0=110{\rm mK} k_B$,  while $T_c \simeq 48 {\rm mK}$.

We emphasize that these numbers
are rather material dependent. For instance,
in an orthoferrite, a canted antiferromagnet,
the effective wall mass is by a factor 
$10^3$ smaller \cite{sloncz} than the value 
obtained from the D\"oring wall mass (\ref{mass}).
Thus tunneling could 
also occur at much larger potential heights 
and  higher crossover temperatures. 

Next, we briefly address the issue of impurities \cite{BLii},
a more detailed account will be given elsewhere. \cite{lossbraun} 
The analysis so far was based on the fact
that  the magnetic field  is constant throughout the sample. 
A single impurity (or similarly an inhomogeneous field)
can be incorporated into the energy 
(\ref{h}) by adding a term  $\kappa a\delta(x-x_0) 
\sin^2 \phi $ where $\kappa$ is of the order
of the anisotropy constant $K_y$.  
Although the impurity is pointlike, it leads to 
an {\it extended}  potential  $U(X)=-\kappa a\;
{\rm sech}^2((X-x_0)/\delta)$ of width $\delta$ 
for the  wall center. Thus even when $\kappa a$ is of the
order of the strength $2V_0$ of the periodic potential,
the impurity potential only leads to a small variation
$(d/\delta) \kappa a$ between pinning sites
separated by $d\ll \delta$. 
This holds also for a random  impurity distribution 
even in the unrealistic case (for YIG) of high
disorder with one impurity per transverse layer. 
Under the action of an external field $H_y$ along the 
easy axis,
which can be much smaller than the coercivity
$H_c$, all wells created by the impurities
can be rendered unstable such that they no longer trap the wall. 
Localization of the wall is then determined by quantum intereference
effects only which we can characterize by
the Anderson localization length. This length, however, is sufficiently 
large  and explicitly given
by $a N_A^2 (\Delta/2V_0)^2\approx 5\cdot 10^4 a$. \cite{lossbraun}

We note that tunneling in periodic pinning potentials allows much higher
crossover temperatures than tunneling out of a single isolated (metastable)
potential. Indeed, in the presence of an external field along the easy-axis
the total energy is $U(X)=-V_0 \, {\rm sech}^2 X/\delta -
2{\cal A} M_0 H_{\rm ext} X$ where
$V_0={3\sqrt{3}\over 2}{\cal A} \delta M_0 H_c$.
The cross-over temperature
and the WKB-exponent  are then given by
$ T_c=2^{3/4}{5\over 18} {g \mu_B\over k_B}
\sqrt{\pi H_c M_0}{\bar \epsilon}^{1/4} $,
${\cal S}_0=2^{3/4} {6\over 5}\hbar N s
\sqrt{H_c / \pi M_0} {\bar \epsilon}^{5/4}$, where  ${\bar \epsilon}=
1-H_{ext}/H$.
For example, for a YIG-sample of $50\,\AA\times 200 \,\AA$
with $H_c=10 {\rm Oe}$, this leads to cross-over temperatures
in the milli-Kelvin range  $0.5\,{\rm mK} < T_c < 1.4\,{\rm mK}$
while the WKB exponent changes in the interval
$0.2 <{\cal S}_0/\hbar<31.1$.

We now turn to a discussion of {\it quantum tunneling
between the two chirality states of a soliton}. We shall obtain 
explicit estimates for the level splitting $\epsilon$
introduced in Eq. (\ref{hsigmax}).
In addition, we shall see that chirality tunneling provides 
a novel scenario for mesoscopic quantum coherence with
one important advantage that both barrier height and 
bias of the double well can be tuned independently 
by external fields. 

Chirality tunneling involves rotation of 
the spins out of the easy-plane 
and  thus cannot be described within the $XY$-approximation 
which we have used so far. To treat this case
we must go back to the full
action (\ref{SE}) and deal with both polar angels $\phi$ and $\theta$. 
The generalization of the wall dynamics to this situation,
in particular, the reduction to the collective coordinate and the 
dissipation due to spin waves, is 
necessarily more involved but still feasible. \cite{lossbraun} 
However, since this generalization is somewhat outside the scope 
of the present work we shall only quote the essential results here 
and give the details in a forthcoming paper. \cite{lossbraun}
For definiteness we concentrate now on ferromagnets where the 
easy-axis anisotropy exceeds the one along the hard-axis, 
i.e. $K_y\gg K_z$; typical examples are bubble materials. \cite{sloncz}
To take advantage of the resulting approximate symmetry 
around the $z$-axis, we represent the magnetization field as
${\bf \Omega}=(\sin\theta \sin\phi, \cos\theta, \sin\theta\cos\phi)$. 
\cite{sloncz} The
Bloch wall is then described
by a rotation of the spins in the  
$xy$-plane about 
the angle $\theta$, and the chirality switching  
by a rotation in the $xz$-plane about the angle $\phi=\pm\pi$. In 
addition, we allow for
an external magnetic field $H_z$ along the hard axis $z$ with 
which one can tune the barrier height that separates the 
two wall chiralities.

Integrating out the $\theta$-fluctuations around the Bloch wall
and restricting ourselves to uniform rotations in $\phi$ (which is
valid \cite{sloncz} if the wall width $\delta$ is less 
than $\sqrt{J/K_z}$),we obtain an effective Langrangian in $\phi(\tau)$,
\begin{eqnarray}
{\cal L}_c={M_c\over2} {\dot \phi}^2 + V(\phi), \\
V=\kappa \cos^2\phi +\eta\cos\phi +\eta^2/4\kappa,
\end{eqnarray}
where $M_c=N_A{S^2 \pi^2 \delta\over 8a^2 K_y}$ is the effective mass
associated with the chirality dynamics, and the parameters 
$\kappa=2\delta N_A K_z$ and
$\eta=g\mu_B SN_A \pi \delta H_z/a$ characterize the barrier 
potential $V$. Defining the anisotropy field by 
$H_a=4aK_z/g\mu_B S\pi$ and noting that the chirality tunnels
between the potential minima defined by 
$\cos\phi_{min}=-H_z/H_a\equiv \nu -1$, we obtain for the level 
splitting $\epsilon$ 
\begin{equation}
\epsilon=4\gamma\omega_c \sqrt{{\cal S}_c/2\pi}\,\,e^{-S_c},
\end{equation}
where $\gamma$ is a numerical constant of order one. The
instanton action ${\cal S}_c$ and frequency 
$\omega_c$ are given by
\begin{eqnarray}
{\cal S}_c=2\pi S (N_A\delta/a)\sqrt{K_z\over K_y}\,\,\nu^{3/2},\\
\omega_c={8 a\over \pi} \sqrt{K_y K_z}\,\,\nu^{1/2}.
\end{eqnarray}
The crossover temperature becomes $T_c=\omega_c/8 k_B$. 
Note the characteristic power dependence on the external control
parameter $\nu=1-H_z/H_a$ with which the chirality
splitting $\epsilon$ can be changed over a large range. 
In the next section we shall also see how a field $H_x$ 
can be used to offset unwanted bias between the potential minima.

We illustrate these results with some typical numbers.
Choosing $N_A \delta/a \sim 10^4$, $K_y/K_z \sim 10$, 
$\nu\sim 10^{-3}$, $aK_z\sim 1 {\rm K} k_B$, and $S=5/2$, we find for
the chirality splitting $\epsilon\approx 5 {\rm mK} k_B$, while the
crossover temperature is $T_c\approx 13 {\rm mK}$. The values for the 
bandwidth $\Delta$ are roughly the same as before.
This shows that
the splitting $\epsilon$ can be made quite large (on the scale of 
$\Delta$) just by tuning the 
external field along the hard axis, while the crossover temperature 
is still reasonably high. Without field, i.e. $\nu=1$, the 
splitting $\epsilon$ is only of non-vanishing value 
if the wall is narrow and/or if $N_A\sim 1$, that 
means if the system is close to being strictly one-dimensional.

\section{Influence of external fields}
\label{extfields}

In this section we  show that 
external fields allow us to control 
the gauge potential $\alpha$. 
In the presence of external fields
the four degenerate Bloch wall configurations 
$\phi_{QC},\theta={\pi\over 2}$  get deformed 
into  new configurations  
$\phi(x), \theta (x)$.
For moving solitons, $\phi(x-X),\theta(x-X)$, the Berry phase 
term (\ref{swzintro}) becomes 
\begin{equation}
{\cal S}_{WZ}  
= - i \tilde \alpha \tilde C \int_0^{\beta} d\tau \dot X,
 \label{xtilde}
\end{equation}
where 
\begin{equation}
\tilde \alpha = {N_A S \over a} \left | \int_{-L/2}^{+L/2} dx\; \phi'
(1-\cos \theta) \right|,
\label{alphatilde}
\end{equation}
and the chirality has been defined as $\tilde C = {\rm sgn} 
\{ \int dx \; \phi' (1 - \cos \theta) \}$ 
with $\phi'=\partial_x \phi$ . 
Note that $\tilde \alpha$ is proportional to 
the area on the unit sphere
between the north pole and
the trajectory which is traced out by a given spin 
upon passing of the Bloch wall (cf. Fig \ref{phase}). 
Since the Bloch wall shape changes in response to an applied
external field, $\tilde \alpha$ will in general differ
from the value $\alpha=N_A S \pi/a$ of 
the Bloch wall $\phi_{QC}$, $\theta={\pi\over2}$.

An external field is taken into account
by adding a Zeeman term  $g\mu_B {\bf B}
\cdot \sum_i {\bf S}_i$  to the spin-Hamiltonian 
${\cal H}$  (\ref{H}).  Correspondingly, the 
total energy $H$ (\ref{h}) is changed  into 
\begin{equation}
\bar H=H+ N_A {\bf h} \cdot \int\!dx \;{\bf \Omega},\quad
{\bf h}=g {\mu_B S\over a} {\bf B}.
\label{Hbar}
\end{equation}
For fields along the easy-axis or the hard-axis,
the  static configurations satisfy the 
Euler-Lagrange equations $\delta \bar H=0$,
\begin{eqnarray}
\FL
&4J \theta'\; \phi' &\cos\theta\; + 
2J \phi'' \sin \theta + 
K_y \sin\theta \sin 2\phi + \nonumber \\
&&+h_x \sin\phi-h_y \cos\phi =0, \nonumber \\
&-2 J \theta'' &+ \sin 2\theta\; [J  \phi'^2 
-K_y\sin^2\phi -K_z]+ \nonumber \\
&&+ h_x\cos\theta \cos\phi+ h_y\cos\theta \sin\phi -h_z \sin\theta=0\; .
\label{LLfield2}
\end{eqnarray}

We first discuss fields along the hard-axis as they have the most
interesting effect on the Berry phase. 
For $h_{x,y}=0$ but $h_z \neq 0$  
all four configurations that emerge from 
$\phi_{QC}$ in (\ref{phiqc})
are still energetically degenerate: 
The invariance of (\ref{H}) under $\pi$-rotations 
around  the $z$-axis (which remains intact for $h_z\neq 0$) implies 
the degeneracy of configurations of 
opposite charge but same chirality. In addition, 
states  of opposite  chirality and charge
are also degenerate since
with  $\phi(x),\theta(x)$ 
also  $-\phi(x),\theta(x)$  solve
(\ref{LLfield2}) (with $h_{x,y}=0$).

In the limit of large hard-axis anisotropy,
$K_z \gg K_y$,  
the possible $\phi$-configurations are  
still  $\phi_{QC}$ given by  (\ref{phiqc})
while
\begin{equation}
\cos\theta=-h_z/ K_z.
\label{costheta}
\end{equation} 
Inserting this into (\ref{alphatilde}) we have 
\begin{equation}
\tilde \alpha = \alpha (1+h_z/K_z),
\label{alphabz}
\end{equation}
which demonstrates that the topological phase 
(\ref{xtilde}) can indeed be tuned by the 
external field. 

For arbitrary values of the ratio $K_z/K_y$
no analytical solution for the soliton structure
can be found. However, we can convince ourselves
that  $\tilde \alpha$ is still field-dependent:
As is verified by inserting $\phi'=\theta'=0$  
and $\phi=\pm\pi/2$ into (\ref{LLfield2}), the spins 
far away from 
the soliton get pulled out of the easy-plane, 
$\cos\theta=-h_z/(2(K_y+K_z))$. Thus in general
$\tilde \alpha$ is different from $\alpha$.

How does the field dependence in (\ref{alphabz}) 
affect the band structure? 
Let us assume that $\sigma = N_A S {d\over a}$ is 
a positive integer, i.e. $\tilde\alpha\equiv\alpha=\pi \sigma/d$
for $h_z=0$.
The dispersion then has the tight-binding form 
(\ref{ekint}) of Fig. \ref{tightbinding}a
and consists of two {\it coinciding} chirality sheets. 
With increasing field $h_z$ the 
sheets of opposite chirality $C=\pm$ get 
separated, each  shifted 
by $\Delta k=|\tilde\alpha - \alpha|$. At an external field
$h_z^0= K_z a/2SN_A d$, this shift becomes 
$\Delta k=\pi/2d$ and the dispersion  shown in 
Fig.  \ref{tightbinding}b is reached. Thus, 
a system with integer $\sigma$ can be 
continuously transformed 
until it reaches half-integer behavior,
and vice versa.\cite{fieldgauge}  Note that this 
behavior is periodic
in the field with period $2 h_z^0$. \cite{uniform}
Moreover, if the field $h_z(t)$ and thus $\tilde\alpha (t)$ depend
on time, it is clear from the effective Hamiltonian (\ref{hsigmax})
that $d{\tilde\alpha}/dt$ plays the role of a
force driving the Bloch wall in positive/negative $x$-direction for
positive/negative chirality. 
Note that this force has its origin in the ``classical" part
of the Berry phase, $\dot\phi\cos\theta$, and therefore can also be deduced 
from the {\it classical} Landau-Lifshitz equation (\ref{eqofmot}). 
[It is somewhat surprising that
this force, as far as we know, has not been discussed in the literature.]

A similar driving effect  is achieved 
by applying an external field $h_y$ along the easy-axis.
Indeed, inserting $\phi_{QC}(x-X)$ of Eq. (\ref{phiqc}) 
into the Zeeman term 
$h_y \int dx \sin \phi_{QC}=-2 h_y N_A  Q X$
we see that a weak magnetic field acts 
like a (classical) force on the soliton center
where $Q$ is the charge of the soliton. It can be seen that
the phase  $\alpha$ remains unaffected by $h_y$.
Note that in analogy to electromagnetism, where $E=-\dot A/c$,
$h_z$ plays the role
of the vector potential $A$ (albeit chirality dependent), 
while $h_y$ corresponds to the
electric field $E$.
Elsewhere we have discussed in detail \cite{BLii,BL} how such 
forces can give rise to Bloch oscillations
of the Bloch wall --- a magnetic analogue of the Josephson 
effect. Similarly, we expect a variety of effects for oscillating fields
such as resonances due to the Wannier-Stark ladders and
related localization effects.
Here we just note that  external fields along
the easy or hard-axis can be used to drive the system through 
the  Brillouin zone.

Finally, we consider an external field $h_x$ along the propagation 
axis. This field lifts the  degeneracy between walls of opposite 
chirality (with $Q$ fixed),
and we find from (\ref{Hbar}),
$2E_C\equiv \bar H [\phi_{Q,C=+}]-\bar H [\phi_{Q,C=-}]=4\pi Q N_A\delta h_x$,
which is simply the effective Zeeman splitting energy of the two
chirality states.
{}From  the exact solutions \cite{thomas} to (\ref{LLfield2})
we see that the phase becomes $\tilde\alpha=\alpha +C\alpha_0(h_x)$,
where $\alpha_0$ vanishes for $h_x\rightarrow 0$.
Hence the relative phase between walls of opposite $C$ remains
$2\alpha$, independent of the field, and the
effective Hamiltonian (\ref{hsigmax})
becomes
\begin{equation}
{\cal H} = {1\over 2M} (p - \alpha  \sigma_z)^2 
+ V(X) + \epsilon \sigma_x +E_C\sigma_z.
\end{equation}
Qualitatively, we see that the last term shifts the dispersion 
sheets of opposite chirality in opposite {\it vertical} direction. 
In the free limit, $V=0$, the eigenvalues 
are $E_{\pm}=(k^2+\alpha^2)/2M \pm[k\alpha(k\alpha+2ME_C)/M^2
+\epsilon^2+E_C^2]^{1/2}$.
Thus, the results of  Sec. \ref{interference} remain basically unchanged
for $E_C \lsim\epsilon$
with the level splitting at $k=0$ becoming now $2\sqrt{\epsilon^2+E_C^2}$.
For $E_C >\epsilon$ tunneling of the chirality (as discussed 
in Sec. \ref{expts}) and hence its alternation 
in the Brillouin zone will get suppressed. 
For instance, if $\epsilon\sim 10$mK this requires
$H_x$ not to exceed $3\cdot 10^{-4}$Oe 
(for the YIG values of Sec.\ref{expts}).
On the other hand, the field $H_x$ provides a useful tool to enhance 
observability of the chirality switching, since it can be used to
offset  unwanted level detuning and to restore the degeneracy of 
the chirality states.

\acknowledgments
We are grateful to A.S. Arrott and A.J. Leggett for many useful
discussions. This work has been supported by the NSERC of Canada 
and the Swiss NSF (H.B.B.).

\appendix
\section{Coherent States and Berry's phase}
\label{appendixa}

In this appendix we discuss the path integrals for 
coherent spin states \cite{klauder} and, in particular, 
the associated  Berry phases. 
We emphasize  single-valuedness of spin states
and the role of admissible gauges since this is of central
importance for the spin parity effects discussed in the main text.
 
A coherent state is the state of minimal uncertainty 
for spin components transverse to the spin-quantization axis. It
is defined as the maximum eigenstate of $S_z$, $|S,M=S\rangle$, 
rotated  into the direction of
the unit vector ${\bf \Omega}=(\sin\theta 
\cos\phi,\sin\theta \sin\phi, \cos \theta)$,
\begin{equation}
|{\bf \Omega} \rangle = e^{-i S_z\phi}  e^{-i S_y\theta} 
e^{-i S_z\chi}   |S,M=S\rangle, 
\label{cs}
\end{equation}
where ${\bf S}$ is  the  spin operator. 
By construction, the coherent state (\ref{cs})
obeys the eigenvalue equation
${\bf S}\cdot {\bf \Omega} |{\bf \Omega}\rangle 
=S |{\bf \Omega}\rangle$
and  is an eigenvector of 
${\bf S}^2$ with eigenvalue $S(S+1)$. 
By use of Wigner's formula \cite{sakurai}, (\ref{cs}) can 
be expressed as 
\begin{eqnarray}
|{\bf \Omega} \rangle =&e&^{-iS\chi} \sum_{M=-S}^S 
{2S \choose S+M}^{1/2} e^{-iM\phi} 
\Bigl( \cos {\theta\over 2} \Bigr)^{S+M}\nonumber\\
&\times&\Bigl( \sin {\theta\over 2} \Bigr)^{S-M}|S,M\rangle.
\label{cs2}
\end{eqnarray}
The  Euler angle $\chi$ has to be fixed by the 
requirement
that the coherent state be {\it single valued} \cite{berry1}
upon  $\phi \to \phi+2\pi n$, $n=0, \pm 1, \dots$. Thus,
$\chi$ is only allowed to take the following  values
\begin{equation}
\chi=(2n+1) \phi, \qquad n=0,\pm 1,\dots.
\label{gauge}
\end{equation}
For the choices $n=-1$ and $n=0$ we shall use the terms 
``north-" and ``south-pole" gauge, respectively. 
Of course, the results obtained in  either of these 
gauges must be physically equivalent.
Note that this requirement 
of single valuedness
has nothing to do with  the transformation  properties 
of $|{\bf \Omega}\rangle$ under {\it active rotations} 
by $2\pi$ which, of course, will always produce a factor
of  $(-1)^{2S}$ irrespective of the choice of $\chi$.

For later use we list a few important 
properties \cite{klauder,einarsson} of the coherent 
states (\ref{cs2}) in the 
north-pole gauge  $\chi=-\phi$.
{}From (\ref{cs2}) it follows that coherent 
states are in general  not orthogonal
\begin{equation}
\langle {\bf \Omega}'|{\bf \Omega} \rangle=
\Bigl (\cos{\theta'\over 2}  \cos{\theta\over 2} 
+\sin {\theta'\over 2}\sin {\theta\over 2} 
e^{i(\phi-\phi')}\Bigr)^{2S},
\label{overlap}
\end{equation}
since ${\bf \Omega}$ may vary continuously 
on the sphere while there are only $2S+1$ 
mutually orthogonal  spin  eigenstates. For infinitesimally
separated angles, the overlap becomes
\begin{equation}
\langle {\bf \Omega}' | {\bf \Omega} \rangle
= 1+ i S \delta\phi\; (\cos\theta -1 ),
\label{infoverlap}
\end{equation}
where $\delta\phi=\phi'-\phi$.
For the south pole parametrization $\chi=\phi$, the 
overlap between infinitesimally separated states 
becomes
\begin{equation}
\langle {\bf \Omega}' | {\bf \Omega} \rangle
= 1+ i S \delta\phi\; (\cos\theta +1 ).
\label{infoverlapsouth}
\end{equation}
Coherent states also 
form an overcomplete set \cite{klauder}
${2S+1\over 4\pi}\int d\Omega \; 
|{\bf \Omega} \rangle \langle{\bf \Omega} | = 1$,
where $d\Omega=d\phi\;d(\cos\theta)$. 
Although the states are not orthogonal, 
the overlap between different states decreases for 
rapidly large $S$ with increasing angle, since
\begin{equation}
|\langle {\bf \Omega}' | {\bf \Omega} \rangle|=
 \Bigl({1\over2} (1+{\bf \Omega'}\cdot {\bf \Omega})\Bigr)^S.
\label{modoverlap}
\end{equation}
In addition we shall make use of the fact that for 
large $S$, we have 
\begin{equation}
\langle {\bf  \Omega'} | {\bf S} |  {\bf  \Omega} \rangle=
\Bigl (   S {\bf \Omega} + {\cal O} \bigl( \sqrt{S} \bigr) \Bigr) 
\langle {\bf \Omega}' | {\bf \Omega} \rangle.
\label{largeS}
\end{equation}
This relation follows from the  exact expressions of 
the spin matrix elements 
and from the fact that 
fluctuations have size ${\cal O}(\sqrt{S})$
since the overlap  (\ref{modoverlap}) decreases as
$\exp\{-S ({\bf \Omega}'- {\bf \Omega})^2/4 \}$.

We derive now a path integral representation 
for the transition amplitude between two 
spin configurations.  To this end,  we represent the 
state vector of the system as a  product of coherent states  
over all lattice sites  $|\{{\bf \Omega}\}\rangle 
=\bigotimes_{i=1}^{N_L} |{\bf \Omega}_i \rangle$. 
Following the usual procedure \cite{klauder}, 
we slice the interval into
$N$ identical  pieces of length $\epsilon$ 
and insert complete sets of  states 
at each lattice site
and  imaginary time step $\tau_n=n \beta/N$,
\begin{eqnarray}
\langle \{{\bf \Omega}_b\} | e^{-\beta {\cal H}} | 
\{{\bf \Omega}_a\} \rangle &=&\left( \prod_{m=1}^{N-1} 
\prod_{i=1}^{N_L}  \int  \tilde d\Omega_i(\tau_m) \right)\nonumber\\
\times\prod_{n=0}^{N-1}
\langle \{&{\bf\Omega}&(\tau_{n+1})\}|
1-\epsilon {\cal H} | \{{\bf\Omega}(\tau_{n})\} \rangle
\label{pathint},
\end{eqnarray}
where $\tilde d \Omega_i =((2S+1)/4\pi) d \Omega_i$
and $ | \{{\bf\Omega}(\tau_{0})\} \rangle= 
| \{{\bf\Omega}_a \} \rangle$, 
 $| \{{\bf\Omega}(\tau_{N})\} \rangle=
 | \{{\bf\Omega}_b \} \rangle$.
In the limit  of large $S$ 
we use (\ref{largeS}) and write
\begin{eqnarray}
\langle \{{\bf\Omega}(\tau_{n+1})\} &|&
1-\epsilon {\cal H} | \{{\bf\Omega}(\tau_{n})\} \rangle=\nonumber\\
 &(&1-\epsilon {\cal H}[S {\bf \Omega}_i(\tau_n)] )
\langle \{{\bf\Omega}(\tau_{n+1})\}|
 \{{\bf\Omega}(\tau_{n})\} \rangle ,
\label{diagH}
\end{eqnarray}
where ${\cal H}[S {\bf \Omega}_i(\tau_n)]$ is the diagonal 
element of the Hamiltonian and follows from (\ref{H}) 
by substituting ${\bf S}_i$ by $S {\bf \Omega}_i(\tau_n)$. 
A ${\cal O}(S^{3/2})$  correction to this diagonal element
has been dropped following standard reasoning. \cite{haldane}
For large $S$, large deviations between coherent
states at adjacent imaginary time steps are 
exponentially suppressed due to (\ref{modoverlap}). 
Therefore the trajectories in imaginary time  become smooth,
and from (\ref{infoverlap}) we obtain for the 
overlap between coherent states at adjacent 
imaginary time steps
\begin{equation}
\langle \{{\bf\Omega}(\tau_{n+1})\}  |
 \{{\bf\Omega}(\tau_{n})\} \rangle\simeq
\prod_{i=1}^{N_L} \Bigl\{ 1- 
i S\;\delta\phi_i(\tau_n) \bigl(1-\cos\theta_i(\tau_n)\bigr)
\Bigr\}, 
\end{equation}
where 
$\delta\phi_i(\tau_n)=
\phi_i(\tau_{n+1})-\phi_i(\tau_n)$.
These overlap terms are of purely 
kinematical origin and contribute to (\ref{pathint}) 
even in the absence of a Hamiltonian.  
It is these terms which are  responsible 
for the distinct behavior of half-odd integral and integral
spins.  Passing to the time-continuum limit $N \to \infty$ 
we obtain,
\begin{eqnarray}
&\langle \{{\bf \Omega}_b\}| e^{-\beta {\cal H}} | 
\{{\bf \Omega}_a\} \rangle
=\left( \prod_{i=1}^{N_L}\int\!
{\cal D}\Omega_i(\tau) \right)\times \nonumber\\
&e^{-\! \int_0^\beta \! d\tau
\left( i S \sum_i  \dot \phi_i (\tau) (1-\cos\theta_i(\tau) )
+ {\cal H} [S  \Omega_i(\tau)]
\right)}, 
\end{eqnarray}
where ${\cal D} \Omega_i(\tau)= \prod_n  \int \tilde d\Omega_i (\tau_n)$ 
is  the measure,
and  we replaced 
$\epsilon \delta \phi_i(\tau)$ by $ d\phi_i(\tau)/d\tau$.

In the space-continuum limit
where the spin configurations  
vary slowly over the lattice constant $a$
the exchange term in ${\cal H}[S\Omega_i]$ 
becomes  $\sum_{i,\rho} {\bf \Omega}_i \cdot 
{\bf \Omega}_{i+\rho}= \int {d^3 r\over a} 
\sum_i (\nabla \Omega_i)^2$.
The transition amplitude then takes finally the form
\begin{equation}
\langle \{{\bf \Omega}_b\} | e^{-\beta {\cal H}} | 
\{{\bf \Omega}_a\} \rangle=
\int \! {\cal D} \phi \; {\cal D} (\cos\theta) e^{-{\cal S}_E[\phi,\theta]},
\label{z3d}
\end{equation}
where the path integral runs over configurations
that satisfy ${\bf \Omega} (x,0)={\bf \Omega}_a(x)$, 
${\bf \Omega}(x,\beta)={\bf\Omega}_b(x)$. The Euclidean 
action is given by ${\cal S}_E={\cal S}_{WZ} + 
\int_0^\beta \! d\tau\; H,$
where the dynamics is determined by the 
Wess-Zumino or Berry phase term
\begin{equation}
{\cal S}_{WZ} = i\;{S\over a^3} \int d^3r  
\int_0^\beta  d\tau \; \dot \phi \; (1-\cos\theta), 
\label{swz}
\end{equation}
and the energy of the spin configuration is given by
\begin{eqnarray}
H &=&  \int {d^3r\over a^3}
\Biggl(
\tilde J S^2 a^2 [ (\nabla\theta)^2 + 
\sin^2\theta (\nabla \phi)^2 ] \\ \nonumber
 &-& \tilde K_y S^2 \sin^2\theta \sin^2\phi + 
\tilde K_z  S^2  \cos^2 \theta \Biggr).
\label{sh}
\end{eqnarray}
Eqs. (\ref{z3d})-(\ref{sh})  {\it generalize 
the formalism of micromagnetics to include 
quantum interference effects}. 
Our discussion is not restricted to the anisotropy 
configurations shown here, one could also include 
e.g. magnetostatic interactions of more general form.
In the particular case where the spin configuration only depends
on one coordinate, we recover (\ref{pathintintro})-
(\ref{h}). 

Note that the Berry phase term (\ref{swz})
has been derived in the  
the north-pole gauge $\chi=-\phi$. If, instead, we had used
the south pole gauge $\chi=\phi$ we would have obtained
\begin{equation}
{\cal S}_{WZ} = -i\;{S\over a^3} \int d^3r  
\int_0^\beta  d\tau \; \dot \phi \; (1+\cos\theta).
\label{swzsouth}
\end{equation}
This gauge dependence can be traced back to 
the gauge dependence of the overlap 
(\ref{infoverlap}), (\ref{infoverlapsouth}) 
of  infinitesimally separated coherent states. 
It is instructive to express this overlap as a line integral 
\begin{equation}
\langle {\bf \Omega}''| {\bf \Omega}'\rangle \simeq e^{
 iS \int_{{\bf \Omega}'}^{{\bf \Omega}''}
d {\bf \Omega}\cdot {\bf A}_{N,S} },
\label{overlapintro}
\end{equation}
over a ``vector potential" 
${\bf A}_{N,S}={\bf e}_\phi(\cos\theta \mp 1)/ \sin\theta $,
where the upper (lower) sign corresponds to the north (south)-pole 
parametrization.
These vector potentials \cite{dirstring} are equivalent to 
that of a  magnetic monopole of unit strength
evaluated on the surrounding unit sphere.  
The gauge character of the different 
parametrizations of the coherent state (\ref{cs}) 
becomes  now apparent.  
If we gauge transform the coherent  state 
$|{\bf \Omega}\rangle \mapsto e^{-i\Lambda} 
|{\bf \Omega}\rangle$, where $\Lambda=\lambda \phi$,
the overlap (\ref{overlapintro}) transforms according 
to ${\bf A}\mapsto {\bf A} + \nabla_\phi \Lambda$.  
By the choice of the gauge 
one decides whether a part of the  Berry phase 
``disappears" in the definition of the
coherent state or whether it appears 
explicitly in the path integral via the overlap
(\ref{overlapintro}).
However, in order to preserve the single-valuedness of the coherent
states---our fundamental postulate---only gauge transformations 
 $\exp \{-i\Lambda\}$ 
are admissible \cite{wu,Lgold} that are single valued 
upon $\phi \to \phi+2 \pi$.  Evidently, this is 
the case for $\Lambda=2 S\phi$ (for all $S$) which relates 
the north- and south-pole parametrization.  
On the other hand, for half-odd integer spin this condition is violated 
for $\Lambda=iS\phi$ which  relates the coherent state with 
$\chi=-\phi$ to the one with  $\chi=0$, but the latter is 
not single valued and thus not an admissible state. 
The corresponding vector potential would be 
${\bf A}_0=-\cot \theta {\bf e}_\phi$ and does
not yield the full Berry phase accumulated in 
a closed circuit: $\oint d{\bf \Omega} \cdot {\bf A}_0$ measures the 
area between the trajectory on the unit sphere 
and the equator while $\oint d{\bf \Omega} \cdot {\bf A}_{N,S}$
measures the area between the trajectory and the north 
or south pole. For trajectories crossing the 
dateline \cite{auerbach}
(this is typically the case if spherical coordinates are chosen 
that are adapted to the symmetry of the Hamiltonian),
the phase factor $\exp \{iS \int d{\bf \Omega} \cdot {\bf A}_0 \}$, 
that results from the ``wrong" choice $\chi=0$  
for the coherent state (\ref{cs}), 
{\it does not} coincide with the Berry phase term, 
$\exp \{iS \int d{\bf \Omega} \cdot {\bf A}_{N,S} \}$,
for half-odd integer spins  and would, e.g., lead to a wrong 
semiclassical  spin-quantization.

\section{Evaluation of the Damping Kernel}
\label{appendixb}

In this Appendix we present the derivation 
of the damping kernel (\ref{Ksum}) starting 
from Eq. (\ref{F}).

In order to evaluate $F$ in (\ref{F}) we first 
complete the square in the exponential. As we are working
only to order ${\cal O} (\dot X^2/c^2)$, it is sufficient
to shift $\varphi$ by $\rho\equiv {1\over 2} 
{\cal G}^{-1} {\cal J}$
since 
\begin{equation}
\varphi\cdot [{\cal G} + {\cal K}] \varphi + {\cal J}\cdot \varphi =
(\varphi+\rho)\cdot  [{\cal G} + {\cal K}] (\varphi+ \rho) 
+ {\cal O} (\dot X^3).
\end{equation}
[${\cal G}, {\cal K}$ are Hermitian].  
Thus, Eq. (\ref{F}) can be rewritten as
\begin{eqnarray}
F[X]&=&\int\! {\cal D} \tilde \varphi \; {\rm det} 
\biggl[{\textstyle \int} dx \bigl\{ 
\phi_0'^2 -\phi_0''[\tilde \varphi -\rho]\bigr\} 
\delta(\tau-\tau')  \biggr] \nonumber \\
&\times&\delta \biggl( {\textstyle \int} \phi_0'[\tilde \varphi -\rho]
\biggr)
e^{-N_A \tilde \varphi\cdot [{\cal G}+{\cal K}] \tilde \varphi} \,\, ,
\label{Ftilde}
\end{eqnarray}
where $\tilde \varphi = \varphi + \rho$.
Eq. (\ref{Ftilde}) can now be considerably  simplified.
First we note that $\int dx \phi_0' \rho \propto \int dx \phi_0' 
{\cal G}^{-1} \phi_0''=0$  due to the parity invariance of ${\cal G}$
and (anti-) symmetry of $\phi_0'$ ($\phi_0''$). 
Thus the $\delta$-function enforces $\tilde \varphi$
to be orthogonal to the zero-mode. 
The Gaussian integrations over $\tilde\varphi$ are 
then well defined and the fluctuations have effective
size ${\cal O} (1/\sqrt{N_A})$. Rescaling 
$\hat \varphi =\sqrt{N_A} \tilde \varphi$
 and making use of the
identity ${\rm det} =\exp {\rm tr} \ln $, we can rewrite
${\rm det} (\delta Q/\delta X)$ as
\begin{equation}
\exp \Bigl\{ {\rm tr} \ln \bigl\{ 1 - {\delta/2 \over \sqrt{N_A}} 
\int \phi_0''\;\hat\varphi + (\delta/2) \int \phi_0'' \rho \bigr\} 
\Bigr\},
\end{equation}
where we used  $\int \phi_0'^2 = 2/\delta$ 
(cf. (\ref{wallenergy})), and where the constant
$\exp\{ {\rm tr} \ln (2/\delta) \}$
has been absorbed into the integration measure. 
The second term under the logarithm can be neglected 
for large $N_A$ and the last term being 
proportional to $\dot X^2$ gives rise 
to a pure mass renormalization. 
Neglecting irrelevant prefactors,
Eq. (\ref{Ftilde}) thus becomes 
\begin{eqnarray}
F[X] &=& e^{-  {\Delta M \over 2}  \int\!d\tau  \dot X^2} 
\int {\cal D} \hat \varphi \; 
\delta \bigl({\textstyle{\int}} \phi_0'\hat \varphi \bigr) \;
e^{-\hat \varphi \cdot [{\cal G}+ {\cal K}] \hat \varphi }  
\nonumber \\
&=& e^{-{\Delta M \over 2}\int\! d\tau  \dot X^2} 
{1\over \sqrt{{\rm det'}({\cal G} 
+{\cal K})}}\, ,
\label{Fdet}
\end{eqnarray}
where the prime on the determinant 
denotes omission of the zero mode
which is enforced by the $\delta$-function,
and $\Delta M = {\cal O}((N_A)^0)$ is a small 
mass renormalization whose exact value is 
not of interest here. In the evaluation of the 
determinant we will encounter
several (ultraviolet) divergent terms which also have the 
form of a  mass renormalization of order ${\cal O}((N_A)^0)$.
All these renormalizations will change the mass $M$
into the experimentally observed 
"dressed" Bloch wall mass $M_{\rm eff}$.
We will thus  drop all these renormalization terms 
and simply replace $M \to M_{\rm eff}$ in the action (\ref{sx}).
 
Moreover, since the SG model is known to be  renormalizable
\cite{zinn} and since 
we are interested only in the long time (infrared) behavior 
there is no need here to 
carry out a systematic renormalization procedure to remove 
the short time divergences.

We now turn to the explicit evaluation of the determinant
in (\ref{Fdet}). We make again use of the identity 
$\ln {\rm det} = {\rm tr} \ln$ and expand the logarithm:
\begin{eqnarray}
&\phantom{+}&{1\over \sqrt{ {\rm det'} ({\cal G} +{\cal K})}}
= e^{ -{1\over 2} {\rm tr'} 
\ln  \left( {\cal G} [1+{\cal G}^{-1}{\cal K}]  \right)} \nonumber \\
&=&{1\over \sqrt {{\rm det'}{\cal G}}}
e^{-{1\over 2} {\rm tr'} 
\left( {\cal G}^{-1} {\cal K} -{1\over 2} {\cal G}^{-1} {\cal K}
 {\cal G}^{-1} {\cal K} + {\cal O} ((\dot X/c)^3)   \right)} \, .
\label{trexpand}
\end{eqnarray}
Since ${\cal K}={\cal O}(\dot X/c)$ this represents 
an expansion in increasing powers of $\dot X/c$.
The factor $[{\rm det'} {\cal G}]^{-1/2}$ is independent
of $\dot X$ and is the partition function of spin 
wave fluctuations around the static Bloch wall. 
The trace in (\ref{trexpand}), $ {\rm tr}(\cdot) = 
\sum_{\bf k} \langle {\bf k} |\cdot| {\bf k} \rangle $,
is evaluated in the basis of eigenfunctions 
of ${\cal G}$,
\begin{equation}
{\cal G} |{\bf k} \rangle = \epsilon_{\bf k} |{\bf k}\rangle,
\qquad \epsilon_{\bf k}=Jk^2 + \kappa \omega^2 + K_y,
\label{eigenvalues}
\end{equation}
where $\kappa=J/c^2$. The anisotropy gap $K_y$ will 
have important consequences for the form of 
the damping kernel below. 
The eigenfunctions factorize into a  space and (imaginary) time part, 
$|{\bf k}\rangle = |\omega \rangle |k\rangle $,
where $\langle\tau|\omega\rangle=e^{i\omega \tau}/\sqrt{\beta}$
with Matsubara frequencies 
$\omega = 2\pi\nu/\beta, \; \nu=0,\pm 1, \dots$.
Since the contribution of the zero-mode $\phi_0'$ 
is explicitly excluded in (\ref{trexpand}), 
we only need the spin wave states \cite{prb1}
\begin{equation}
\langle x| k\rangle = N_k [-ik\delta + 
{\rm tanh} (\textstyle{{x\over \delta}})] e^{ikx},
\label{psi}
\end{equation}
where $N_k=[L(1+k^2\delta^2)]^{-1/2}$. 
The $k$ values in (\ref{psi}) are implicitly defined by 
\begin{equation}
kL + \Delta(k) = 2\pi n\, ,
\label{period}
\end{equation}
where $\Delta(k) = 2 \arctan {1\over k\delta}$
is the scattering phase shift of the eigenfunction (\ref{psi}).

To render the results finite in the thermodynamic limit, 
we have to subtract
the vacuum fluctuations \cite{sakita} and thus renormalize,
\begin{equation}
{1\over {\rm det'} ({\cal G} +{\cal K} )} \to 
{{\rm det} ({\cal G}_0 +{\cal K} ) \over 
{\rm det'} ({\cal G} +{\cal K} )},
\label{renorm}
\end{equation}
where ${\cal G}_0=-\kappa \partial_\tau^2 -J\partial_x^2 + K_y $
is the operator describing spin waves around the anisotropy minimum
in the absence of a Bloch wall. 
${\cal G}_0$ has the {\it same} eigenvalues (\ref{eigenvalues}) 
as ${\cal G}$ but the space eigenfunctions are 
simply plane waves where the $k$ values are 
given by $k_{\rm free}=2\pi n/L$ rather than (\ref{period}). 
For the results given below which only involve one summation over $k$,
the renormalization (\ref{renorm}) then simply amounts to the 
replacement 
\begin{eqnarray}
\sum_k  \to \sum_k - \sum_{k_{\rm free}}&=&
\int_{-\infty}^\infty dk [ \rho(k)-\rho_{\rm free}]\nonumber \\
&=&\int_{-\infty}^\infty {dk\over 2\pi} {d\Delta \over dk}\, , 
\label{sumrenorm}
\end{eqnarray}
where $\rho=dn/dk={L\over 2\pi}-{1\over 2\pi}{d \Delta\over dk}$ 
is the density of states corresponding to 
(\ref{period}), $\rho_{\rm free}= 2\pi/L$, and where
we go over now to the thermodynamic limit.
{}From the definition of $\Delta$ it follows that 
$d \Delta /dk =-2\delta/(k^2\delta^2+1)$.

With these preliminaries, we can now rewrite 
the lowest order term in  (\ref{trexpand}) as follows, 
\begin{eqnarray}
-{1\over 2} {\rm tr} {\cal G}^{-1} {\cal K} 
&=&-{1\over 2}  \sum_{\bf k} {\kappa \over \epsilon_{\bf k}}
\bigl[ 2 \langle k | -i\partial_x|k\rangle \;\langle \omega| i 
\dot X \partial_\tau | \omega \rangle \nonumber \\
&-& \langle k |\partial_x^2|k\rangle 
\langle \omega | \dot X^2  | \omega \rangle
\bigr],
\end{eqnarray}
where $\kappa=J/c^2$.
Using the eigenfunctions (\ref{psi}) we obtain 
$ \langle k| -i\partial_x | k'\rangle =  k \delta_{k k'} 
+ {\cal O}(L^{-1}) $ and  $ \langle k|\partial_x^2 | k'\rangle = 
- k^2 \delta_{k k'} + {\cal O}(L^{-1}) $.
Inserting the identity $1=\int d\tau |\tau \rangle \langle \tau|$ 
we obtain in leading order in $L$, 
\begin{equation}
-{1\over 2} {\rm tr} {\cal G}^{-1}  {\cal K}
={\kappa \over 2\beta} \sum_{\bf k}
  {k \over \epsilon_{\bf k}} \int d\tau \Bigl\{2\omega\dot X - k\dot X^2
\Bigr\}.
\end{equation}
The first term on the rhs vanishes 
since $\epsilon_{\bf k}$ is symmetric in both 
$k$ and $\omega$ and the second term leads to a 
mass renormalization which diverges logarithmically 
(after the partial renormalization  (\ref{sumrenorm})). 
As mentioned above, this term is part of the  
dressing  of the  ``bare" D\"oring mass to the experimentally
observed value $M_{\rm eff}$, and thus there is no need to remove
this divergence explicitly.

The damping due to spin waves will be exclusively due to 
the remaining terms in (\ref{trexpand}) which shall be discussed  next.
Using the above notation, we have up to order $\dot X^2$
\begin{eqnarray}
{1\over 4} {\rm tr} \left({\cal G}^{-1} {\cal K}\right)^2  
&=& \kappa^2 \sum_{{\bf k},{\bf k}'} 
{1\over \epsilon_{\bf k} \epsilon_{{\bf k}'}}
|\langle {\bf k} | \dot X \partial_\tau \partial_x | {\bf k}'\rangle |^2.
\label{trace2}
\end{eqnarray}
In leading order in $L$ we have
\begin{equation}
\langle  {\bf k} | \dot X \partial_\tau \partial_x | {\bf k}'\rangle =
-{k\omega'\over \beta} \delta_{kk'}
 \int\!d\tau \;  e^{i(\omega'-\omega) \tau} \dot X(\tau).
\end{equation}
Thus Eq. (\ref{trace2}) can be rewritten in the form
\begin{equation}
{1\over 4} {\rm tr} \left({\cal G}^{-1} {\cal K}\right)^2
= \int_0^\beta \!\!\!d\tau  \int_0^\beta \!\!\!d\sigma
\dot X(\tau)\;\dot X(\sigma)\;\gamma(\tau-\sigma) ,
\label{damp1}
\end{equation}
with 
\begin{equation}
\gamma(\tau) = {1\over \beta^2}  \sum_{\omega,\omega',k}
{ k^2 \omega \omega' e^{i(\omega'-\omega) \tau}
\over 
[\omega^2 + \omega_k^2]
[\omega'^2 +\omega_k^2] },
\label{gamma}
\end{equation}
where $\omega_k^2=c^2(k^2 + \delta^{-2})$.
With partial integrations 
and with $\gamma(\tau+\beta)=\gamma(\tau)$  
Eq. (\ref{damp1}) reduces to
\begin{equation}
{1\over 4} {\rm tr} \left({\cal G}^{-1} {\cal K} \right)^2 
= -{1\over 2}\!\int_0^\beta \!\! d\tau \!\!\int_0^\tau \!\! d\sigma\,  
K(\tau-\sigma)\;[ X(\tau) -  X(\sigma)]^2 , 
\label{damp2}
\end{equation}
where $ K(\tau) = - 2 \partial_\tau^2 \gamma $.
In (\ref{damp2}),   we have neglected a term 
$2 [X(\beta) - X(0)]\int d\tau \dot X \gamma$
which turns out to be small for typical 
tunneling processes.  For the 
evaluation of $\gamma$ and $K$ we make use of the 
exact relation 
\begin{equation}
D_\omega(\tau)={2\omega\over \beta} \sum_{n=-\infty}^\infty 
{e^{i\omega_n \tau}
\over \omega_n^2 + \omega^2} = {\cosh\bigl( \omega(|\tau|- 
{\beta\over 2})\bigr)
\over \sinh\left({\beta  \omega \over 2}\right) },
\label{D}
\end{equation}
where $\omega_n=2\pi n/\beta $ and the rhs is understood to 
be periodically extended beyond $|\tau|\leq \beta/2$.
With (\ref{gamma}) and (\ref{D}) we finally obtain 
for $K=-2\partial_\tau^2 \gamma$
\begin{equation}
K(\tau) = \sum_k k^2 \omega_k^2\Bigl[ \sinh^{-2}
\left({\beta\omega_k\over 2}\right)- 2D^2_{\omega_k} (\tau) \Bigr].
\label{K}
\end{equation}
Note that as a consequence of the relevant coupling between the 
system $\dot X$ and bath which is quadratic in the bath coordinates
$\varphi$, $K$ is proportional to $D^2_\omega$ rather than 
$D_\omega$ as in the usual Caldeira-Leggett theory. 
For low temperatures, the damping kernel (\ref{K}) reduces to 
$K(\tau)=-2 \sum_k k^2 \omega_k^2 e^{-2\omega_k|\tau|}$.

\begin{figure}
%\epsfxsize=6cm
%\leavevmode
%\epsfbox{figure1.ps}
\caption{ 
The Berry phase factor for one single spin  ${\bf S}$,
$\exp\{ iS \oint d\phi 
(1-\cos\theta)\}=e^{iS A}$, 
where $A$ is the area  on the unit 
sphere enclosed by the trajectory ${\cal C}$ traced 
out by ${\bf S}$.
}
\label{phase}
\end{figure}

\begin{figure}
%\epsfxsize=8cm
%\leavevmode
%\epsfbox{figure2.ps}
\caption{ 
a) Bloch wall configuration with $Q=1$, $C=-1$ 
in a thin long slab centered at the pinning site $X=0$; b)  
Periodic pinning potential $V$ for the wall center $X$.
}
\label{sample} 
\end{figure}

\begin{figure}
%\epsfxsize=8cm
%\leavevmode
%\epsfbox{figure3.ps}
\caption{
Dispersion of a soliton in a weak 
periodic potential.
a) For $\sigma=N_A S d/a$ integer
the dispersion resembles  that of a Bloch electron
and the gaps at $\pm \pi/d$ are due to the 
periodic pinning potential (\protect{\ref{V}}). 
b) For  $\sigma$ half-odd integer
the Brillouin zone is halved and two 
subsequent band minima have opposite chirality.
Band gaps $2\epsilon$ arise due to 
tunneling between the chiralities as described by 
(\protect{\ref{hsigmax}}), $E_+$ and $E_-$ are 
the dispersions as given by (\protect{\ref{epm}}).
}
\label{freedisp}
\end{figure}

\begin{figure}
%\epsfxsize=8cm
%\leavevmode
%\epsfbox{figure4.ps}
\caption{
Soliton dispersion in 
the tight binding limit. 
a) For $\sigma=N_A S d/a$ integer
a standard tight binding dispersion
results. 
b) For $\sigma$ half-odd integer, the 
Brillouin zone (and bandwidth) is halved and
two subsequent band minima belong to 
opposite chiralities. A gap $2\epsilon$
develops if the two chiralities of
the soliton are connected by tunneling. 
}
\label{tightbinding}
\end{figure}


\begin {references}
\bibitem{footnote:general}The treatment of the general
case is more involved but still possible; the results are similar in 
nature and will be presented elsewhere. \cite{lossbraun}
\bibitem{gap}
The lowest energy
spin wave modes  are the Winter modes
\protect{\cite{winter,janak}}
with dispersion $\epsilon_{\bf k} =
(2\sqrt{J}a/S)k\sqrt{J {\bf k}^2 + K_z}$ where
${\bf k} = (k_y,k_z)$ is the wavevector
transverse to the sample axis. These modes are
gapless for an infinite sample but they acquire a finite-size gap
for a mesoscopic sample since
the minimal wavevector component is $k_{{\rm min}}=\pi/w$, where
$w$ is the maximal transverse width.
Thus, the spin waves are frozen out for
$k_B T<\epsilon_{\bf k_{{\rm min}}}$,
or equivalently, if at temperature $T$ the sample width
is smaller than
$w_0(T)= \pi \sqrt{2J/K_z}[\sqrt{1+\tau^2}-1]^{-1/2}$
with $\tau=k_B T S/K_z a$. Note that $w_0$
increases with decreasing temperature. E.g., for YIG
with parameters as in Sec. \protect{\ref{expts}},
we have $w_0=400\AA$ at $T=0.5\;{\rm K}$, while
at $T=50\;{\rm mK}$ we have $w_0=2800\AA$.
\bibitem{spinhalf} Note that on-site anisotropies are 
ineffective for $S=1/2$. On the other hand, exchange 
anisotropy supports the existence of ferromagnetic 
solitons also in this case 
(see e.g. Ref. \protect{\onlinecite{mikeska}}),
and thus Eq. (\protect{\ref{SE}}) may be regarded as an effective 
model for such soliton excitations even for $S=1/2$.
\bibitem{compactness} At this point we  reduce 
the topology of  the 2-sphere $S^2$
to the $S^1$ topology of the spin confined to the  easy-plane. 
In Sec. \protect{\ref{interference4}} we shall consider 
what happens if the soliton can tunnel between 
two chiralities, which belong to two distinct 
topological sectors of the sine-Gordon field
configuration.
\bibitem{spinwaves}
In micromagnetic notation this reads 
$\omega=2 (\gamma/M_0) ([Ak^2+K_e+K_h][Ak^2+K_e])^{1/2}$,
where $A,K_e,K_h$ are defined as in Sec. \protect{\ref{model}},
and $\gamma/M_0=a^3/\hbar S$.
\bibitem{breathers} Note, however, that high energy soliton-soliton 
breather solutions of Ref. \protect{\onlinecite{breath}}
do not have a counterpart in the sine-Gordon model.
\bibitem{lambda}
This notation differs from Ref. \onlinecite{LDG}.
Denoting the (positive) anisotropy constants of the 
latter work with  $\bar K_y$, $\bar K_z$,
we have $\bar K_y=(a/NS^2)K_y$ and 
$\bar K_z=(a/NS^2)(K_y+K_z)$.
\bibitem{zeromode} Note that the boundary 
conditions on the field $\phi$
(which contains a Bloch wall) ensure that $Q$ has 
indeed a zero:  Since $\phi_0'(x)$ 
is a localized positive function, the sign of $Q$ will
be proportional to the sign of $\phi$ which 
changes at least once as a function of $x$ due
to the boundary conditions.
\bibitem{pinning}
The strength $V_0$ of the 
pinning potential can be related to 
a classical coercivity via $V_0/{\cal A}=H_c M_0 d/\pi$
where ${\cal A}$ is the sample cross sectional area. 
$H_c$ is defined as the field $H$ at which the force 
$2{\cal A} M_0 H$ on the domain wall 
due to the external field renders the 
potential $V(X)$ unstable.
\bibitem{footnote2walls} In a finite system,
strictly speaking, periodic
boundary conditions correspond physically to a ring geometry
which would require a pair of $\pi$-Bloch walls.
However, in the thermodynamic limit
the Bloch walls can  be separated far enough from each other
so that we may treat them independently.
\bibitem{footnotepc} Note that this is quite
different to, e.g., a persistent current, where the 
counterpart of $\alpha$, the Aharonov-Bohm flux,
scales inversely proportional to the sample length
at constant field. See also Sec. \protect{\ref{interference4}}.
\bibitem{footnote:halving} The period halving is  
directly seen if, instead, we choose vanishing 
boundary conditions. In this case we may perform a
gauge transformation on (\protect{\ref{heff}})
with $e^{-i\alpha X \sigma_z}{\cal H}e^{i\alpha X \sigma_z}
=p^2/2M + \epsilon \sigma_x \cos(2\alpha X)+
\epsilon \sigma_y \sin(2\alpha X) + V(X)$, which leaves the partition
function unaltered. Thus, the induced space potential has now
double the period, $2d$ (with interchanged chiralities),
which translates into half the period
in reciprocal space. Note that this argument cannot
be used for periodic boundary conditions, since the 
wave functions in the new gauge are no longer periodic.
\bibitem{coercivity} This value of the coercivity is an empirical value
since no data are known for sufficiently small YIG-samples
at the temperatures of interest. For macroscopic samples,
coercivities as low as $10^{-2} {\rm Oe}$ are reported. 
\protect{\cite{dillon}} 
Since the coercivity is expected to increase with decreasing sample 
size and temperature, we assume a coercivity of $2 {\rm Oe}$.
\bibitem{fieldgauge}
This effect is  again gauge independent. 
In an arbitrary gauge (\protect{\ref{gauge}}) we have
$\tilde \alpha_n=(\sigma/d) 
\int d\phi([2n+1]-\cos\theta)=\tilde\alpha+
n C \sigma{2\pi\over d}$ since $\phi_{QC}$ flips the spins
by $C\pi$. Thus, the positive and negative
chirality branch are each shifted by a 
reciprocal lattice vector (or half a reciprocal 
lattice vector for $\sigma$ half-integer).
Hence for integer and half-integer 
$\sigma$ the spectra obtained in different gauges
merely differ by multiples of 
a reciprocal lattice vector $2\pi/d$
and thus are identical.
\end{references}
\end{document}